\documentclass[]{jfm} % lineno to number lines

\usepackage{graphicx}
\usepackage{newtxtext}
\usepackage{newtxmath}
\usepackage{natbib}
\usepackage{hyperref}
\hypersetup{
    colorlinks = true,
    urlcolor   = blue,
    citecolor  = blue,
    allcolors  = blue
}

\newcommand{\RomanNumeralCaps}[1]
\linenumbers

\usepackage{xcolor}
\usepackage{subcaption}

\newcommand{\derp}[2]{\partial #1/ \partial #2}

\newcommand{\Derp}[2]{\frac{ \partial #1 }{ \partial #2 }}

\newcommand{\derm}[2]{{D #1 }/{ D #2 } }
\newcommand{\Derm}[2]{\frac {D #1 } {D #2 } }

\newcommand {\mb}[1]{\mathbf{#1}}

\newcommand{\qhat}{\hat{\mb{q}}}

\newcommand{\Malt}{\check{M}}

\newcommand{\ubar}{\bar{u}}
\newcommand{\rhobar}{\bar{\rho}}
\newcommand{\mubar}{\bar{\mu}}
\newcommand{\Tbar}{\bar{T}}

\newcommand{\lbar}{\bar{\lambda}}
\newcommand{\barxi}{\bar{\xi}}

\newcommand{\gl}{\gamma_\lambda}
\newcommand{\gmu}{\gamma_\mu}

\newcommand{\rhoT}{\bar{\rho}_{T}}

\newcommand{\phirho}{\partial_\rho \Phi}
\newcommand{\phiT}{\partial_T \Phi}
\newcommand{\phirhobar}{\partial_\rho \bar{\Phi}}
\newcommand{\phiTbar}{\partial_T \bar{\Phi}}
\newcommand{\thetabar}{\bar{\theta}}

\newcommand{\COt}{$\text{CO}_2$ }

\usepackage{esint} % closed integral 
\newcommand{\nablaa}{\nabla}
\newcommand{\divop}{\nabla \cdot}

\newcommand{\ugradp}{(\bb{u} \cdot \nabla)}

\newcommand{\dd}{\mathrm{d}}

\newcommand{\bb}[1]{\textbf{#1}}

\newcommand{\bos}[1]{\boldsymbol{#1}}
\definecolor{persoBlue}{HTML}{0076c4} %{B33009}
\definecolor{persoGreen}{HTML}{06ab00} %{B33009}
\definecolor{persoRed}{HTML}{fc1303} %{B33009}
\definecolor{persoDarkPurple}{HTML}{730382} %{B33009}
\definecolor{blueresponse}{RGB}{2, 109, 209}
\definecolor{greenresponse}{RGB}{26, 138, 61}

\newcommand{\rev}[1]{#1}
\newcommand{\revv}[1]{#1}

\shorttitle{Linear stability of viscous stratified shear flows with arbitrary equations of state}
\shortauthor{B. Bugeat}

\title{Linear stability of viscous stratified shear flows with arbitrary equations of state}

\author{B. Bugeat\aff{1} 
  \corresp{\email{bb283@le.ac.uk}}
}

\affiliation{\aff{1}School of Engineering, University of Leicester, Leicester, United Kingdom
}
\begin{document}
\maketitle

\begin{abstract}
A mathematical framework is developed to analyse the linear stability of viscous stratified flows with arbitrary equations of state. 
The model accommodates equations of state of the form $p=\Phi(T,\rho)$ or $\rho=\phi(T)$, together with a set of governing dimensionless parameters. 
Density stratification is unrestricted, allowing for the study of non-Boussinesq effects. 
A low-Mach-number approximation is employed to filter out acoustic phenomena.
The framework is applied to two viscous stratified configurations of plane Couette flow: a triple-layer density profile supporting the Taylor–Caulfield instability (TCI), and a differentially heated supercritical fluid exhibiting a distinct instability mechanism. 
Following validation against reference studies, we present new results showing that non-Boussinesq effects weaken the viscous TCI, whilst viscosity can further destabilise the supercritical fluid instability, in contrast to the stabilising influence of buoyancy.
\end{abstract}

\begin{keywords}
% Instability, Stratified flows, Taylor-Caulfield Instability, Supercritical Fluid, non-Boussinesq, Low-Mach Approximation
\end{keywords}

%%%%%%%%%%%%%%%%%%%%%%%%%%%%%%%%%%%%%%%%%%%%%%%%%%%%%%%%%%%%%%%%%%%%%%%%%%%%%
%%%%%%%%%%%%%%%%%%%%%%%%%%%%%%%%%%%%%%%%%%%%%%%%%%%%%%%%%%%%%%%%%%%%%%%%%%%%%
\section{Introduction}
Stratified flows, broadly defined as flows in which density or other fluid properties vary, arise in a wide range of natural and industrial settings, including oceanic and atmospheric dynamics, energy systems and chemical engineering. Such flows support a rich variety of hydrodynamic instabilities.
Their linear stability has classically been analysed using the Taylor-Goldstein equation \citep{taylor1931effect,goldstein1931stability}, derived for inviscid shear flows with density stratification.
\rev{It has been successfully used in a wide array of geophysical problems and has led to fundamental results such as the Miles-Howard theorem, which states that two-dimensional, linear normal modes are stable if $Ri>1/4$ is verified everywhere in the flow \citep{miles1961stability,howard1961note}, where $Ri$ is the Richardson number.}
One limitation of the Taylor-Goldstein equation is the Boussinesq approximation on which it relies, which accounts for density variations in the gravity term of the momentum equation but neglects them in the inertial terms \citep{vallis2017atmospheric}.
It also assumes a linear equation of state (EoS) between density and temperature.
These assumptions are valid in shear flows provided that density variations are small ($\Delta \rho / \rho \ll 1$); see \cite{guha2018inertial} for a deeper discussion, including a special result for the limiting case of weak shear.

For larger density variations, the Boussinesq approximation may fail to capture important physical mechanisms.
First, even in low-Mach-number flows, the velocity field may no longer be divergence-free since a fluid parcel may significantly expand due to (diffusive) heat transfer.
This occurs when $\alpha_p \Delta T$ is no longer small, with $\alpha_p$ the isobaric thermal expansion coefficient and $\Delta T$ a characteristic temperature difference in the flow \citep{mayeli2021buoyancy,barletta2022boussinesq}.
Second, when the pressure and density gradients are not collinear, the baroclinic torque contributes to the production of vorticity.
This has been shown to affect, for example, the growth of the Kelvin-Helmholtz instability in variable-density mixing layers \citep{soteriou1995effects,fontane2008stability} and in compressible jets \citep{lesshafft2007linear}.
The properties of the Holmboe instability \citep{holmboe1962behavior} are also altered by this mechanism \citep{umurhan2007holmboe,barros2011holmboe}.
A final example, which is of interest in this paper, is the Taylor-Caulfield instability (TCI) \citep{taylor1931effect,caulfield1995experimental} which occurs in shear flows with multiple density interfaces.
\cite{heifetz2015stratified} showed that the TCI is weakened by non-Boussinesq effects, as the baroclinic torque slows down the internal gravity waves travelling along the density interfaces, whose interaction gives rise to this instability; see \cite{carpenter2011instability} for a review of the interacting-wave interpretation of hydrodynamic instabilities. 

Non-Boussinesq effects also play a major role in the dynamics of supercritical fluids (SCFs), whose pressure and temperature are above the thermodynamic critical point.
In the region of the Widom line \citep{banuti2015crossing}, relatively small temperature differences can induce large fluid property variations (in particular, $\Delta \rho / \rho \simeq 1$).
As a result, strong, localised gradients of fluid properties are observed in SCF flows.
The linear stability of such flows has been increasingly studied in recent years, in various configurations: mixing layers \citep{ly2022,wang2022linear}, boundary layers \citep{ren2019boundary,bugeat2022new}, channel flows \citep{ren2019linear,bernades2025linear} or Couette flows  \citep{bugeat2024inst}.
A striking result is the existence of a hydrodynamic instability specific to SCFs, which was first predicted by \cite{ren2019boundary}.
The authors named this instability `mode II' in order to distinguish it from the unstable Tollmien-Schlichting \revv{(TS)} waves coexisting in a boundary layer flow; in the present paper, we will refer to it as the SCF instability, since other fluids and flows will also be considered.
The baroclinic torque is one of the main ingredients causing this instability, which can be interpreted within the interacting-wave framework mentioned earlier.
The steep, localised mean gradients of density and viscosity support the propagation of two vorticity waves generated by shear and baroclinic effects, whose interference in turn produces two self-amplifying interacting waves \citep{bugeat2024inst}.

Whilst buoyancy is known to affect the dynamics and heat transfer of turbulent flows of SCFs \citep{yoo2013turbulent}, its effect on the stability and transition to turbulence in these fluids has received less attention.
Observations in recent experiments of horizontal channel flow have shown that stable stratification could delay transition to turbulence \citep{draskic2025stability}.
One linear stability analysis, carried out in a boundary layer flow, showed that the SCF instability is indeed weakened by stable stratification \citep{boldini2026first}. 
Further efforts are needed to understand these effects at a more theoretical level.

Most of the stability analyses for SCFs cited above were carried out using the linearised, fully compressible Navier-Stokes equations (with complex equations of state).
Whilst SCFs find applications in the compressible regime \citep{guardone2024nonideal}, \rev{many applications occur at low Mach numbers.
The speed of sound in supercritical \COt is of the order of $10^2 \> \mathrm{m.s^{-1}}$ \citep{zevnik2006ultrasound}, whereas internal flows, e.g. in heat exchangers, operate at much lower velocities.
For example, in the experimental channel flow of \cite{whitaker2024flow}, the bulk velocity is of the order of $1 \> \mathrm{m.s^{-1}}$ (estimated from the published parameters).
A similar estimation for the natural circulation loop of \citep{gao2023numerical} yields a characteristic velocity of approximately $0.1 \> \mathrm{m.s^{-1}}$. 
This corresponds to Mach numbers of the order of $10^{-3}-10^{-2}$.}
Moreover, the mechanism of the SCF instability mentioned above is incompressible in nature: the large density variations initiating the baroclinic torque arise from temperature variations rather than pressure effects.
Modelling the stability and transition to turbulence of SCFs at low Mach numbers using the fully compressible Navier-Stokes is therefore often unnecessarily complex.
It makes the implementation of the model more difficult and error-prone as it contains a large number of terms---in SCFs, fluid properties vary with both temperature and pressure.
Moreover, when numerically advancing the equations in time, the (fast) acoustic timescale must be resolved; this greatly restricts the magnitude of the time step, potentially increasing computational cost by orders of magnitude.
Finally, it may obscure the interpretation of the results because of the presence of acoustic modes that are irrelevant to the hydrodynamic stability.

The aim of the present paper is to formulate a mathematical framework for the linear stability analysis of viscous stratified flows with an arbitrary EoS.
This framework embeds non-Boussinesq effects, without restriction on the shape or magnitude of density variations, accounting in particular for baroclinic torque and non-linear equations of state.
Variations in fluid properties (viscosity, heat capacity, thermal conductivity) are allowed, although the case of constant coefficients will also be considered and discussed.
As compressible effects are often negligible in such flows, we seek a simpler model than the linearised compressible Navier-Stokes equations.
To this end, we adopt a low-Mach number asymptotic formulation, in which flow variables are expanded in a power series of the Mach number squared \citep{paolucci1982filtering}.
Building on the extension for real gases proposed by \cite{battista2014turbulent}, we derive a linear stability model that bridges the gap between Boussinesq formulations and fully compressible approaches. 
The resulting framework retains the essential physics associated with strong property variations whilst avoiding the complexity of compressible models.
In the case of constant fluid coefficients and linear EoS, we will see that the present framework is not significantly more complex than the linear viscous Boussinesq equations \citep{eaves2017multiple,parente2020modal,thummar2024stability} whilst remaining applicable to a much broader class of flows at low Mach numbers.
In addition to supercritical fluids, this model could for example be applied to liquids with non-linear EoS \citep{olsthoorn2019diffused,bhavsar2026effects,estay2026}, dense gases \citep{gloerfelt2020dense}, or heating and ventilation systems with large temperature variations (e.g. the mixed convection over a heated plate).

As shown later, one limitation of the proposed model is its validity only for flows whose vertical height is much smaller than a characteristic height $L^*$ above which the hydrostatic pressure gradient affects the thermodynamic pressure.
For an ideal-gas atmosphere, $L^*$ is the classical scale height $r^*T^*/g^*$ (with $r^*$ the specific gas constant, $T^*$ the temperature, and $g^*$ the gravitational acceleration), which is of the order of $10^4 \> m$ for the Earth's atmosphere.
Therefore, the present framework does not apply to deep atmospheric flows. 
Instead, it is well suited for flows with strong density variations that extend on a height scale smaller than $L^*$.
We note that the present framework is different from the anelastic approximation, which is also based on the idea of filtering acoustic waves.
The anelastic approximation was specifically developed in the context of atmospheric flows, relying on the assumption of constant potential temperature \citep{vallis2017atmospheric}.
Contrary to the present model, it allows the hydrostatic pressure gradient to affect the background pressure.
Another key difference from the proposed framework lies in the formulation of the continuity equation, which, in particular, neglects the time derivative of the density field.

The derivation of the non-linear equations in the limit of small Mach numbers and for arbitrary equations of state is first presented in section \ref{sec.theory}, with particular attention paid to the choice of dimensionless quantities.
The case of a fluid with constant properties and a linear EoS between temperature and density is discussed, highlighting the links with the Boussinesq approximation.
After introducing convenient notations that facilitate the implementation of the model, the general linearised equations governing stability are presented (section \ref{sec.linear_dynamics}).
In section \ref{sec.results}, the proposed framework is applied to two cases: the TCI and the SCF instability.
\rev{These applications serve to rigorously validate the proposed approach whilst demonstrating its versatility in modelling different classes of stratified flows. 
We also highlight the potential of the framework for future studies by presenting new results on the influence of non-Boussinesq effects, viscosity, and buoyancy on these instabilities.
These effects are discussed in terms of disturbance vorticity and wave interaction.}

%%%%%%%%%%%%%%%%%%%%%%%%%%%%%%%%%%%%%%
\section{Theory} \label{sec.theory}

\subsection{Navier-Stokes equations for arbitrary equations of state} \label{sec.NL}
In dimensional form, the continuity, momentum and energy equations read
\begin{align}
  \Derm{\rho^*}{t^*} &= - \rho^*  \divop \bb{u}^* \\
  \rho^* \Derm{\bb{u}^*}{t^*} &= - \nablaa{p^*} +  \divop{\bos{\tau}^*} + \rho^* \bb{g}^* \\
  \rho^* \Derm{h^*}{t^*} &=  \Derm{p^*}{t^*} + \nabla \cdot \left( \lambda^* \nabla T^* \right) + \bos{\tau}^* \cdot \nabla \bb{u}^* ,
\end{align}
\rev{where superscript $^*$ denotes dimensional variables.}
$\derm{}{t} = \derp{}{t} + \ugradp$ is the material derivative and $\bb{u}$ is the velocity field.
The pressure $p$, temperature $T$, and density $\rho$ are related by an equation of state formulated as
\begin{align}
  p^* = \Phi^* (T^*,\rho^*) ,
\end{align}
which is the natural form of many thermodynamic models (e.g. ideal gas, van der Waals or Redlich–Kwong).
The vector \bb{g} is the gravitational acceleration and $\bos{\tau}$ is the viscous stress tensor.
The dynamic viscosity $\mu$ and thermal conductivity $\lambda$ may depend on both $T$ and $\rho$.
The enthalpy $h$ verifies
\begin{align}
  \dd h^* = c_p^* \dd T^* + \rev{ \partial_{p^*} h^* } \dd p^* ,
\end{align}
\rev{where $c_p^*$ is the specific heat capacity at constant pressure and $\partial_{p^*} h^*$ the partial derivative of $h^*$ with respect to $p^*$.
Using $\partial_p h^*= (1 - \alpha_p^* T^*) / \rho^*$, where $\alpha_p^*$} is the isobaric thermal expansion coefficient, the energy equation can be rewritten
\begin{align}
  \rho^* c_p^* \Derm{T^*}{t^*} &=  \alpha_p^* T^* \Derm{p^*}{t^*} + \nabla \cdot \left( \lambda^* \nabla T^* \right) + \bos{\tau}^* \cdot \nabla \bb{u}^* .
\end{align}
To nondimensionalise these equations, a reference pressure $p_0^*$ is chosen together with a reference temperature $T_0^*$.
The reference density $\rho_0^*$ is then obtained from the EoS.
The dynamic viscosity $\mu_0^*$, the thermal conductivity $\lambda_0^*$, the specific heat capacity $c_{p,0}^*$ and the thermal expansion coefficient $\alpha_{p,0}^*$ are also introduced, corresponding to the fluid properties at ($T_0^*$, $\rho_0^*$).
Finally, a characteristic velocity $u_0^*$ and a length scale $d^*$ are chosen.  
The dimensionless equations are then found as
\begin{align}
  \Derm{\rho}{t} &= -\rho \divop{\bb{u}} , \\
  \rho \Derm{\bb{u}}{t}  &= - \frac{1}{\check{M}^2} \nablaa{p} + \frac{1}{Re} \divop{\bos{\tau}} - \frac{1}{Fr^2} \rho \bb{e}_y , \\
  \rho c_p \Derm{T}{t} &=  \tilde{A} \Psi \alpha_p T  \Derm{p}{t} + \frac{\Psi \check{M}^2}{Re} \bos{\tau} \cdot \nabla \bb{u} + \frac{1}{Pr Re} \nabla \cdot \left( \lambda \nabla T \right) ,
\end{align}
where $\bb{e}_y$ is an upward-pointing unit vector.
The following dimensionless numbers have been introduced
\begin{align}\label{eq.NDnumer}
  \Malt = \frac{u_0^*}{\sqrt{p_0^*/\rho_0^*}} \> \> , \> \>  Re = \frac{u_0^* d^*}{\nu_0^*} \> \> , \> \> Fr = \frac{u_0^*}{\sqrt{g^* d^*}}  \> \> , \> \> Pr = \frac{\nu_0^*}{\kappa_0^*} \> \> , \> \> \Psi = \frac{p_0^*}{\rho_0 c_{p,0}^* T_0^*} \> \> , \> \> \tilde{A} = \alpha_{p,0}^* T_0^* ,
\end{align}
where $\nu_0^* = \mu_0^* / \rho_0^*$ and $\kappa_0^* = \lambda_0^* / (\rho_0^* c_{p,0}^*)$ are the kinematic viscosity and the thermal diffusivity at ($T_0^*$,$p_0^*$), respectively.
$Re$, $Fr$ and $Pr$ are the Reynolds, Froude and Prandtl numbers, respectively.
Following \citet{battista2014turbulent}, we have introduced a pseudo Mach number, $\Malt$.
For an ideal gas, $\Malt = \sqrt{\gamma} M$, where $M$ is the classical Mach number and $\gamma$ is the heat capacity ratio.

We will make the important assumption that $(\Malt/Fr)^2 \ll 1$, which can be interpreted by noticing that $(\Malt/Fr)^2 = \rho_0^* g^* d^* / p_0^*$.
The numerator is the characteristic hydrostatic pressure difference developing over a height $d^*$.
The assumption $(\Malt/Fr)^2 \ll 1$ is verified when this pressure difference is negligible compared with $p_0^*$, i.e. when hydrostatic pressure gradients do not affect the thermodynamic pressure profile.
For flows shallow enough to verify this condition, we will find that the thermodynamic pressure is therefore uniform (equation \ref{eq.p0}).
For deeper flows that may not verify $(\Malt/Fr)^2 \ll 1$, the gravity term of the momentum equations should be retained, leading to a (hydrostatic) variation of pressure at leading order.
For an ideal gas, we note that $(\Malt/Fr)^2 = d^* / L^*$, where $L^* = r^* T_0^* / g^*$.
For air, this is the atmospheric scale height, which is of the order of $10^4 \> m$.

We have also introduced two dimensionless groups, $\Psi$ and $\tilde{A}$, which, like $Pr$, only depend on the fluid properties at a given thermodynamic state.
These two numbers carry little physical meaning in the present framework: they are simply assumed finite and will disappear from the energy equation following the asymptotic expansion carried out in section \ref{sec.lowMapprox}.
Values of $\Psi$ and  $\tilde{A}$ are reported for air, water, and supercritical \COt in appendix~\ref{app.Psi_Atilde}.
Further insight into $\Psi$ can be gained by considering an ideal gas, for which it is related to the thermodynamically meaningful dimensionless quantity $\gamma$, the heat capacity ratio, through $\Psi = (\gamma - 1) / \gamma$.
We note that, therefore, $0<\Psi < 1$.
For a real gas, $\Psi$ can be expressed in terms of two classical dimensionless numbers: $\Psi = Z_0 / (C_{p,0}^*/R^*)$.
Here, $Z_0$ is the compressibility factor at the reference thermodynamic state and reads $Z_0 = p_0^* / (\rho_0^* r^* T_0^*)$, where $r^*$ is the specific gas constant.
The other dimensionless number, $C_{p,0}^*/R^*$, is defined from $C_{p,0}^*$, the molar heat capacity, and $R^*$ the universal gas constant.
Similarly, the dimensionless number $\tilde{A}$ does not have an immediate interpretation because of the presence of the absolute temperature scale, $T_0^*$, used in the non-dimensionalisation procedure.
The flow temperature difference $\Delta T^*$ will later be introduced through two dimensionless parameters, $A = \alpha_p^* \Delta T^*$ and $\Lambda = \Delta T^* / T_0^*$, in sections \ref{sec.constantcoeff} and \ref{sec.results}, respectively.
$A$ controls the magnitude of the thermal expansion whilst $\Lambda$ is the dimensionless flow temperature difference.
$\tilde{A}$ is related to these parameters according to $\tilde{A} = A / \Lambda$.

\subsection{Approximation for small Mach numbers}\label{sec.lowMapprox}
Assuming that $\Malt \ll 1$, we expand all variables in a power series in $\Malt^2$ \citep{paolucci1982filtering}.
For example, the pressure is written 
\begin{align}
  p = p^{(0)} + \Malt^2 p^{(2)} + \mathcal{O}(\Malt^4) .
\end{align} 
At leading order, the momentum equation yields
\begin{align}\label{eq.p0}
  \nablaa p^{(0)} = 0 
\end{align}
\revv{where we have used the `shallow-stratified-flow' assumption $(\Malt/Fr)^2 \ll 1$, discussed in the previous section.}
At the next order, we have
\begin{align}
  \rho^{(0)} \Derm{\bb{u}^{(0)}}{t}  &= - \nablaa p^{(2)} + \frac{1}{Re} \divop{\bos{\tau}^{(0)}} - \frac{1}{Fr^2} \rho^{(0)} \bb{e}_y 
\end{align}
The dimensionless EoS is, at leading order,
\begin{align}
  p^{(0)} = \Phi(T^{(0)},\rho^{(0)}) 
\end{align}
This leads to the classical interpretation of $p^{(0)}$ as the thermodynamic pressure and $p^{(2)}$ as the hydrodynamic pressure.
For an open system, $p^{(0)}$ is time-independent \citep{nicoud2000conservative}.
Combined with \eqref{eq.p0}, we deduce that $\derm{p^{(0)}}{t} = 0$, and $p^{(0)}$ is therefore constant, uniform, and equal to the ambient pressure $p_0$, i.e. the reference pressure.
As a result, the dimensionless EoS for an open system in the limit of small Mach numbers reads
\begin{align}
  \Phi(T^{(0)},\rho^{(0)}) = 1 ,
\end{align}
The energy equation is, at leading order,
\begin{align} \label{eq.energy}
  \rho^{(0)} c_p^{(0)} \Derm{T^{(0)}}{t} &= \frac{1}{Pr Re} \nabla \cdot \left( \lambda^{(0)} \nabla T^{(0)} \right)
\end{align}
Since $\derm{p^{(0)}}{t} = 0$, we have $\derm{\rho^{(0)}}{t} = \rho_T^{(0)} \derm{T^{(0)}}{t}$, where we have introduced
\begin{align}
  \rho_T = \left. \Derp{\rho}{T} \right|_p .
\end{align}
This quantity can be deduced from the EoS $\Phi$, as shown in the next subsection.
The continuity equation can therefore be written
\begin{align} \label{eq.divU1}
  \divop{\bb{u}^{(0)}} = - \frac{\rho_T^{(0)}}{\rho^{(0)} }  \Derm{T^{(0)}}{t}  .
\end{align}
We note that this equation could alternatively be formulated in terms of $\alpha_p$; however, $\rho_T$, which also appears in several other parts of the framework, is preferred in order to reduce the overall number of variables.
Equation \eqref{eq.divU1} can also be recast using the energy equation \eqref{eq.energy} as 
\begin{align} \label{eq.divU2}
  \divop{\bb{u}^{(0)}} = - \frac{1}{Pr Re} \frac{\rho_T^{(0)}}{[\rho^{(0)}]^2 c_p^{(0)}}  \nabla \cdot \left( \lambda^{(0)} \nabla T^{(0)} \right) ,
\end{align}
showing that the volume change of a fluid parcel is purely driven by heat transfer in the limit of low-Mach numbers; there are no compressibility effects associated with pressure.
We note, in particular, that $\divop{\bb{u}} = 0$ for inviscid flows ($Re \to \infty$), for which the diffusion of both momentum and energy becomes negligible. 
In this case, a fluid parcel is advected at constant temperature, and thus its volume remains constant.

\subsubsection{General case: summary of the set of equations}\label{sec.general_case_summary_of_the_set_of_equations}
The final set of non-linear equations is summarised below, where, without ambiguity, we have dropped the notations associated with the order of the power series; as previously mentioned, the pressure $p$ is purely hydrodynamic and is now decoupled from the EoS.
\begin{align}
  \divop{\bb{u}} &= - \frac{\rho_T}{\rho} \Derm{T}{t} \label{eq.NL1} , \\
  \rho \Derm{\bb{u}}{t}  &= -  \nablaa{p} + \frac{1}{Re} \divop{\bos{\tau}} - \frac{1}{Fr^2} \rho \bb{e}_y \label{eq.NL2} , \\
  \rho c_p \Derm{T}{t} &= \frac{1}{Pr Re} \nabla \cdot \left( \lambda \nabla T \right) \label{eq.NL3} , \\
  \Phi(T,\rho) &= 1 \label{eq.phi1}
\end{align}
Importantly, $\rho_T$ is obtained from the EoS as 
\begin{align}
  \rho_T =   \rev{ - \frac{\partial_T \Phi}{\partial_\rho \Phi} } .
\end{align}
This model can therefore be applied to various equations of state by prescribing the function $\Phi$; for completeness, we also discuss the case of an implicit EoS in appendix \ref{app.imp}.
\rev{The partial derivatives $\phiT$ and $\phirho$} can be obtained either analytically or numerically (see section \ref{sec.disturbance_fluidprop}).
Finally, a caloric law for $c_p$, together with viscosity and thermal conductivity laws, can be introduced to account for property variations with $T$ and $\rho$.

\subsubsection{Special case 1: the equation of state does not depend on the pressure}\label{sec.eosPhi}
If the EoS does not depend on pressure, such that
\begin{align} \label{eq.phiT}
  \rho = \phi(T) 
\end{align} 
replaces equation \eqref{eq.phi1}, then the set of equations (\ref{eq.NL1}-\ref{eq.NL3}) can directly be used, with $\rho_T$ now readily obtained from \eqref{eq.phiT}.

\subsubsection{Special case 2: constant coefficients with linear equation of state} \label{sec.constantcoeff}
We consider the case where $c_p$, $\lambda$, and $\mu$ are constant and where the dimensional EoS is linear, given by
\begin{align}\label{eq.eoslinear}
  \rho^* = \rho_0^* \left[ 1 - \rev{ \alpha_{p,0}^*} (T^* - T_0^*) \right] ,
\end{align} 
In this  case, it is more natural to introduce the dimensionless temperature $\theta$ as
\begin{align}
  \theta &= \frac{T^*- T_0^*}{\Delta T^*}
\end{align}
where $\Delta T^*$ is a characteristic temperature difference of the problem.
Note that this dimensionless temperature could have been introduced in the general model (\ref{eq.NL1}-\ref{eq.phi1}), but was not preferred since a temperature scale based on an absolute temperature is more suitable for thermodynamic relations.
The following set of equations is obtained:
\begin{align}
  \divop{\bb{u}} &= \frac{A}{\rev{\rho}} \Derm{\theta}{t} \label{eq.NLconst1}, \\
  \rho \Derm{\bb{u}}{t}  &= -  \nablaa{\tilde{p}} + \frac{1}{Re} \divop{\bos{\tau}} + Ri  \theta \bb{e}_y \label{eq.NLconst2} , \\
  \rho \Derm{\theta}{t} &= \frac{1}{Pr Re}  \nabla^2 \theta  \label{eq.NLconst3},
\end{align}
where the dimensionless hydrostatic pressure has been absorbed in $\tilde{p} = p + y / Fr^2$.
We have introduced the dimensionless number 
\begin{align}
  A = \alpha_{p,0}^* \Delta T^* ,
\end{align} 
whilst the Richardson number naturally appeared in the equations as
\begin{align}\label{eq.richardson}
  Ri = \frac{\alpha_{p,0}^* \Delta T^* g^* d^*}{(u_0^*)^2} .
\end{align}
We note that $Fr$, $Ri$ and $A$ are not independent, as $Fr^2 = A / Ri$.
The set of equations (\ref{eq.NLconst1}-\ref{eq.NLconst3}) holds strong similarities with the viscous Boussinesq approximation \citep{eaves2017multiple,parente2020modal,thummar2024stability}.
We highlight two major differences, \rev{which are well-known in non-Boussinesq frameworks.}
First, the present framework allows for a non-zero divergence of the velocity when thermal expansion becomes significant \citep{barletta2022boussinesq}; this effect is controlled by the dimensionless number $A$.
Second, density variations are retained in the inertial term of the momentum equation. 
This introduces an inertial contribution to the baroclinic torque, which is absent in the Boussinesq approximation \citep{guha2018inertial} (see also the vorticity equation in section \ref{sec.vorticity_eq}).
\rev{The Boussinesq approximation is recovered when $A \ll 1$, which leads to $\divop{\bb{u}} = 0$ at leading order (equation \ref{eq.NLconst1}).
We also verified that, in this case, density variations do not affect the inertial terms of the momentum equation, since $\rho  = 1$ at leading order (as readily found from $\rho = 1 - A \theta$, the dimensionless form of equation \ref{eq.eoslinear})}.

\subsection{Linear dynamics} \label{sec.linear_dynamics}
The linearised equations governing the dynamics are obtained by adding small perturbations to a baseflow, defined as a steady solution of the non-linear equations derived in section \ref{sec.NL}.
The flow variables can be grouped into the state vector $\bb{q} = (u, v, w, p, T)^T$, where {$\bb{u} = (u,v,w)^T$}.
The baseflow is denoted $\bar{\bb{q}}$.
Using Cartesian coordinates $(x,y,z)$, the flow is assumed to be parallel, such that {$\bar{\bb{u}} = (\bar{u}(y), 0 , 0)^T$}. 
Similarly, the other baseflow variables only depend on $y$ (e.g. $\bar{\rho}(y)$, $\bar{T}(y)$, $\bar{c}_p(y)$, etc.).
Gravity is assumed to act along the $y$-axis.
The disturbances are decomposed in normal modes as
\begin{align} \label{eq.mode}
  \hat{\bb{q}}(x,y,z,t) = \bb{q} (y) e^{i (\alpha x + \beta z - \omega t)} + c.c.
\end{align} 
where c.c. stands for complex conjugate.
In a temporal stability analysis, the streamwise and spanwise wavenumbers $(\alpha,\beta) \in \mathbb{R}^2$ are parameters of the linear model, whilst $\omega \in \mathbb{C}$ is the model output.
An instability is identified when the imaginary part of $\omega$ is positive for at least one mode.

\subsubsection{Disturbance fluid properties} \label{sec.disturbance_fluidprop} 
Since $\Phi(\rho,T)=1$ for all $\rho$ and $T$ (equation \ref{eq.phi1}), it follows that {$\Phi(\bar{T}, \bar{\rho}) = \Phi(\bar{T} + T, \bar{\rho}+\rho) = 1 $}.
Therefore, the first-order Taylor expansion of {$\Phi(\bar{T} + T, \bar{\rho}+\rho)$} yields
\begin{align}\label{eq.rhoT1}
  \rho = \bar{\rho}_T T ,
\end{align} 
where $\bar{\rho}_T = \rev{ - \phiTbar  / \phirhobar}$, the overbar indicating that the functions are evaluated at the temperature and density of the baseflow $(\Tbar, \rhobar)$.
We now introduce a set of notations that will allow for a more compact form of the linearised equations.
Let us consider the disturbance viscosity as an illustrative example, which is readily found as 
\begin{align}\label{eq.mu}
  \mu = \rev{ ( \partial_T \bar{\mu} ) T + ( \partial_\rho \bar{\mu} ) \rho ,}
\end{align}
where $\partial_T \bar{\mu}$ and $\partial_\rho \bar{\mu}$ are the partial derivatives of $\mu$ at $(\Tbar, \rhobar)$.
Using \eqref{eq.rhoT1}, we can recast \eqref{eq.mu} as a function of $T$ only.
We find that 
\begin{align} \label{eq.mu2}
  \mu = \gamma_\mu T , 
\end{align}
where we have introduced
\begin{align} \label{eq.gamma}
    \gamma_\mu &=  \rev{ \partial_T \bar{\mu} + \bar{\rho}_T \partial_\rho \bar{\mu}    . }
\end{align}
Note that $\gamma_{\mu}$ depends on the baseflow profile and can be obtained prior to the linear analysis.
An analogous quantity can be calculated for $\lambda$, which will be noted $\gamma_\lambda$.

Finally, we note that the partial derivatives $\partial_T \bar{f}$ and $\partial_\rho \bar{f}$ of any function $f$ evaluated at $(\bar{T},\bar{\rho})$ can be numerically obtained using a finite-difference approximation such as
\begin{align}\label{eq.FD}
  \rev{\partial_T \bar{f}} = \frac{f(\bar{T} + \varepsilon, \bar{\rho}) - f(\bar{T} - \varepsilon, \bar{\rho})}{2 \varepsilon} \> \> , \> \> \rev{ \partial_\rho \bar{f} } = \frac{f(\bar{T} , \bar{\rho} + \varepsilon) - f(\bar{T} , \bar{\rho} - \varepsilon)}{2 \varepsilon}
\end{align}
where $\varepsilon \ll 1$ is a numerical parameter to be chosen.
Results should not depend on its value; in practice, $\varepsilon = 10^{-6}$ is found adequate in a dimensionless framework (see \cite{mettot2014computation} for a more extensive discussion).
This approach has the advantage to be systematic; it circumvents the need to hardcode each thermodynamic derivative ($\phiTbar$, $\phirhobar$, $\partial_\rho \mubar$, etc.), allowing for a quick implementation of new thermodynamic and diffusion laws.

\subsubsection{Linear equations: general framework} \label{sec.AEoSmodel}
The equations of the linear dynamics of viscous stratified flow in the limit of small Mach numbers for an arbitrary EoS are ultimately found as
\begin{align}
  i \alpha u + v' + i \beta w &= - \frac{\rhoT}{\bar{\rho}} \left[ i ( \ubar \alpha - \omega ) T + \Tbar' v  \right] \label{eq.L1} \\
  i \rhobar ( \alpha \ubar - \omega ) u + \rhobar \ubar' v &= - i \alpha p + \frac{1}{Re} \mathcal{V}_x \label{eq.L2} , \\
  i \rhobar ( \alpha \ubar - \omega ) v  &= - p'   + \frac{1}{Re} \mathcal{V}_y - \frac{1}{Fr^2} \rhobar_T T \label{eq.L3}, \\
  i \rhobar ( \alpha \ubar - \omega ) w  &= - i \beta p   + \frac{1}{Re} \mathcal{V}_z \label{eq.L4}, \\
  \rhobar \bar{c}_p \left[ i ( \ubar \alpha - \omega ) T + \Tbar' v  \right]  &= \frac{1}{Pr Re} \mathcal{E} \label{eq.L5},
\end{align}
where the prime notation denotes differentiation with respect to $y$.
The diffusion terms $\mathcal{V}_x$, $\mathcal{V}_y$, $\mathcal{V}_z$ and $\mathcal{E}$ are given below for a Newtonian fluid with varying viscosity and thermal conductivity, using the notations $\gamma_\mu$ and $\gamma_\lambda$ introduced in section \ref{sec.disturbance_fluidprop}:
\begin{align}
  \mathcal{V}_x &= - ( \frac{4}{3} \alpha^2 + \beta^2 ) \mubar u + \mubar' u' + \mubar u'' + i \alpha \mubar' v + \frac{1}{3} i \alpha \mubar v' - \frac{1}{3} \alpha \beta \mubar w + (\ubar'' \gmu + \ubar' \gmu') T + \ubar' \gmu T'  \\
  \mathcal{V}_y &= - (\alpha^2 + \beta^2 ) \mubar v + \frac{4}{3} \mubar' v' + \frac{4}{3} \mubar v'' - \frac{2}{3} i \alpha \mubar' u + \frac{1}{3} i \alpha \mubar u' - \frac{2}{3} i \beta \mubar' w + \frac{1}{3} i \beta \mubar w' + i \alpha \ubar' \gmu T  \\
  \mathcal{V}_z &= - ( \alpha^2 + \frac{4}{3}  \beta^2 ) \mubar w + \mubar' w' + \mubar w''- \frac{1}{3} \alpha \beta \mubar u  + i \beta \mubar' v + \frac{1}{3} i \beta \mubar v' , \\
  \mathcal{E} &= - ( \alpha^2 + \beta^2 ) \lbar T +  ( \Tbar'' \gl + \Tbar' \gl') T + (\lbar ' + \Tbar' \gl ) T' + \lbar T'' .
\end{align}

\subsubsection{The case of constant coefficients and linear equation of state} \label{sec.linearEoSmodel}
The linearised equations in the case of constant coefficients and linear EoS presented in section \ref{sec.constantcoeff} read, for a Newtonian fluid,
\begin{align}
  i \alpha u + v' + i \beta w &= \frac{A}{\rev{\rhobar}} \left[ i ( \ubar \alpha - \omega ) \theta + \thetabar' v  \right] \label{eq.L1const} \\
  i \rhobar ( \alpha \ubar - \omega ) u + \rhobar \ubar' v &= - i \alpha p + \frac{1}{Re} \left[ -(\alpha^2 + \beta^2) u + u'' + \frac{1}{3} ( - \alpha^2 u + i \alpha v' - \alpha \beta w )  \right] \label{eq.L2const} , \\
  i \rhobar ( \alpha \ubar - \omega ) v  &= - p' + \frac{1}{Re} \left[ -(\alpha^2 + \beta^2) v + v'' + \frac{1}{3} ( i \alpha u' +  v'' + i \beta  w' )  \right] + Ri  \theta \label{eq.L3const}, \\
  i \rhobar ( \alpha \ubar - \omega ) w  &= - i \beta p  + \frac{1}{Re} \left[ -(\alpha^2 + \beta^2) w + w'' + \frac{1}{3} ( - \alpha \beta u +  i \beta v' - \beta^2  w )  \right] \label{eq.L4const}, \\
  i \rhobar ( \ubar \alpha - \omega ) \theta + \rhobar \thetabar' v    &= \frac{1}{Pr Re} \left[ -(\alpha^2 +  \beta^2) \theta + \theta'' \right] \label{eq.L5const}.
\end{align}

\subsection{Vorticity equation}\label{sec.vorticity_eq}
The different physical mechanisms contributing to the linear dynamics can be identified from the disturbance vorticity equation.
For simplicity, only two-dimensional perturbations ($\beta=0$) are considered.
The equations are presented in physical space to facilitate the interpretation of the individual terms.
Combining equations \eqref{eq.L2} and \eqref{eq.L3}, the equation of the disturbance vorticity $\xi = \derp{v}{x} - \derp{u}{y}$ reads, in the general model,
\begin{equation} \label{eq.linear_vorticity}
    \Derp{\xi}{t} + \bar{u} \Derp{\xi}{x} = \> \> \underbrace{- \> \> v \barxi'  }_{S_\xi} \> \> \underbrace{ - \> \> \frac{\bar{\rho}'}{\bar{\rho}^2}  \Derp{p}{x}}_{B_\xi} 
    \> \> \underbrace{ - \> \> \frac{1}{Fr^2} \frac{\rhobar_T}{\rhobar} \Derp{T}{x} }_{G_\xi}
    \> \> \underbrace{ - \> \> \barxi \divop \bb{u} }_{E_\xi} 
    \> \> + \> \> V_\xi ,
    \end{equation}
where $\barxi = - \ubar'$ is the \revv{baseflow} vorticity and $V_\xi$ contains all the viscous terms, reading
\begin{equation} \label{eq.linear_vorticity_const}
    V_\xi =   \frac{1}{Re} \left[ \frac{1}{\rhobar} \right( \Derp{V_y}{x} - V_x' \left) + \frac{\rhobar'}{\rhobar^2} V_x \right] .
\end{equation}
$S_\xi$ represents the vertical transport of  vorticity by the perturbations.
$B_\xi$ is the (non-Boussinesq) baroclinic torque, which is at play when the pressure and density gradients are non-collinear.
This term arises from density variations in the inertial terms of the momentum equation, which are neglected in the Boussinesq approximation.
$G_\xi$ is the vorticity generation by buoyancy.
It is also referred to as the gravitational baroclinic torque \citep{guha2018inertial}, as it arises from the misalignment of the horizontal temperature (or density) gradient and the gravitational acceleration vector (i.e. the vertical hydrostatic pressure gradient).
$E_\xi$ is associated with the thermal expansion or contraction of the fluid.

In the case of constant coefficients and linear EoS, the 2D disturbance vorticity equation can be obtained combining equations \eqref{eq.L2const} and \eqref{eq.L3const}.
It contains the same terms as equation \eqref{eq.linear_vorticity}, but the buoyancy term now reads
\begin{equation} \label{eq.linear_vorticity_const}
    G_\xi^{(\mathrm{const})} =  \frac{Ri}{\rhobar} \Derp{\theta}{x} .
\end{equation}
Moreover, the viscous term can be simplified to
\begin{equation} \label{eq.linear_vorticity_const}
    V_\xi^{(\mathrm{const})} = \frac{1}{Re} \left[ \frac{1}{\rhobar} \nabla^2 \xi + \frac{\rhobar'}{\rhobar^2} \left( \nabla^2 u + \frac{1}{3} \Derp{}{x} \divop \bb{u} \right) \right] .
\end{equation}

\subsection{Numerical approach}
After choosing a suitable discretisation method---the pseudospectral method of \citep{orszag1971accurate} based on Chebyshev polynomials is used in this work---the system of equations is recast as
\begin{align}
  - i \omega \mathbb{B} \qhat =  \mathbb{A} \qhat .
\end{align}
The stability of the system is determined by solving this eigenvalue problem, with the imaginary part $\omega_i$ being the growth rate of a mode; its positivity indicates an instability.
The MATLAB function \textit{eig} is used to compute the eigenvalues and corresponding eigenvectors.
Plane Couette flow, bounded by two walls, is considered in this work.
Dirichlet boundary conditions are imposed at the walls for all velocity components (no-slip condition) as well as for temperature (isothermal condition).
These conditions are enforced in the momentum and energy equations of matrix $\mathbb{A}$ whilst the corresponding rows of matrix $\mathbb{B}$ are set to zero.
The number of collocation points, $N$, required to achieve mesh-independent results may vary depending on the physical cases and parameter values.
A convergence study is presented in appendix \ref{app.meshconv}.
In the cases explored in this work, strong gradients arise not only near the walls but also in the central region of the domain.
Since Chebyshev collocation points, $y_i \in [-1,1]$, are clustered near the walls, the bijection defined below is used to remap them to a new set of points, $\hat{y}_i \in [-1,1]$, that is more evenly distributed across the flow:
\begin{align}
   \hat{y}_i = \frac{y_i^3 + a y_i}{1 + a} ,
\end{align}
where $a=0.2$.
It has been verified that the implementation of the mapping did not have any effect on the results, whilst reducing the number of collocation points required for convergence and, consequently, the computational cost.

%%%%%%%%%%%%%%%%%%%%%%%%%%%%%%%%%%%%%%
\section{Validation and Applications}\label{sec.results}
\begin{figure}
    \centering    
    \includegraphics[angle=-0,trim=0 0 0 0, clip,width=0.995\textwidth]{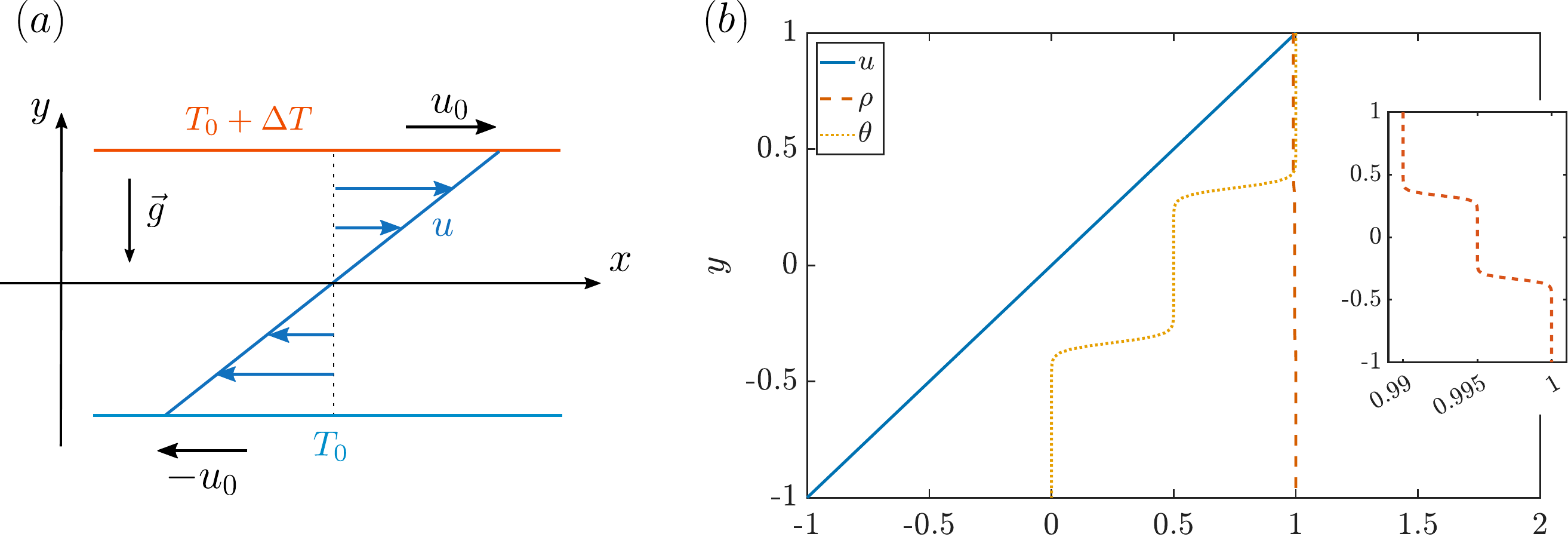}
    \captionsetup{justification=justified}
    \caption{(a) Sketch of plane Couette flow with plates maintained at different temperatures.
    \rev{(b) Triple-density-layer baseflow considered in section \ref{sec.TCI}, shown for $A=10^{-2}$.
    The inset provides a close-up of the density profile.}
    }
    \label{fig.sketch}      
\end{figure}

The mathematical framework developed in the previous section is applied to two physical cases mentioned in the introduction: the TCI and the SCF instability.
The purpose of this section is twofold. 
In each case, we first validate our model against data from the literature.
We then move beyond the validation case to demonstrate how additional effects can be examined with the proposed framework.
In both cases, plane Couette flow is considered with $-u_0^*$ and $+u_0^*$ at the bottom and top plates, respectively (figure \ref{fig.sketch}a).
A temperature gradient is imposed by maintaining the bottom and top plate at temperatures $T_0^*$ and $T_0^* + \Delta T^*$, respectively.
We introduce the dimensionless temperature difference
\begin{align}
  \rev{\Lambda} = \frac{\Delta T^*_0}{T_0^*} .
\end{align}
Only stable stratifications will be considered ($\rev{\Lambda}>0$).
The half-gap distance $d^*$ is used as a characteristic length scale.
All the following stability analyses are two-dimensional ($\beta=0$).

\subsection{Taylor-Caulfield instability} \label{sec.TCI}

\subsubsection{Validation}
We consider the viscous Taylor-Caulfield instability studied by \cite{eaves2017multiple} within the viscous Boussinesq approximation.
This instability develops in a three-layer density-stratified baseflow with constant shear, and has zero phase velocity.
Our aim is to recover the results of the maximum growth rate over a range of Reynolds numbers, Richardson numbers, and disturbance wavelengths (figure 3 in \cite{eaves2017multiple}). 
We use the model with constant coefficients and linear EoS presented in sections \ref{sec.constantcoeff} and \ref{sec.linearEoSmodel}.
Given our choice of dimensionless variables, the three-layer baseflow density profile considered in the reference study translates into
\begin{align}\label{eq.rhoEaves}
  \rhobar(y) = 1 - \frac{A}{2} \left\{ 1 +  \frac{1}{2}  \left[  \tanh R \left( y - \frac{1}{3} \right) + \tanh R \left( y + \frac{1}{3} \right)   \right]  \right\},
\end{align}
where $R$ is a dimensionless parameter that controls the thickness of the diffused interfaces between each density layer, located at $y=-1/3$ and $y=1/3$.
Note that the parameter $A$ in equation \eqref{eq.rhoEaves} is not present in the Boussinesq model of \cite{eaves2017multiple} and that the definition of our Richardson number, $Ri$ (equation \ref{eq.richardson}), is different from theirs, referred to as $Ri_{E}$.
This is because the density difference used to define this dimensionless number differs by a factor of 2 between the two studies: we use the difference between the two plates ($\rho_0^* \alpha^* \Delta T^*$) whilst they consider the difference between one plate and the centre line.
Moreover, the reference density is taken at a different location: the bottom plate in the present work, and the centre line in the reference paper. 
As a result, the relation between the two Richardson numbers is found as ${Ri = 2 Ri_{E} ( 1 - A/2 )}$.
Note that $A$ is strictly lower than 1; otherwise, density would be negative.
In the Boussinesq limit ($A \ll 1$, see section \ref{sec.constantcoeff}), we have $Ri = 2 Ri_{E}$. 

The density profile of equation \eqref{eq.rhoEaves} is, in our framework, implemented via the baseflow temperature profile ${\thetabar(y) = (1 -  \rhobar(y)  ) / A}$, which is a reformulation of equation \eqref{eq.eoslinear}.
The baseflow velocity is trivially $\ubar=y$.
\rev{The baseflow profile is shown in figure \ref{fig.sketch}(b).}
Following the reference study, we set $R=30$ and $Pr=1$.
A small value of $A$ is used in order to approach the Boussinesq regime; we use $A = 10^{-4}$.
No-slip and isothermal ($T=0)$ boundary conditions are used for the disturbances.
Our calculations recover the maximum growth rates of \cite{eaves2017multiple} well, as shown in figure \ref{fig.tripleDensityLayer}(a).

\subsubsection{Non-Boussinesq effects}\label{sec.TCI_varyA}
\begin{figure}
    \centering
    \includegraphics[angle=-0,trim=0 0 0 0, clip,width=0.99\textwidth]{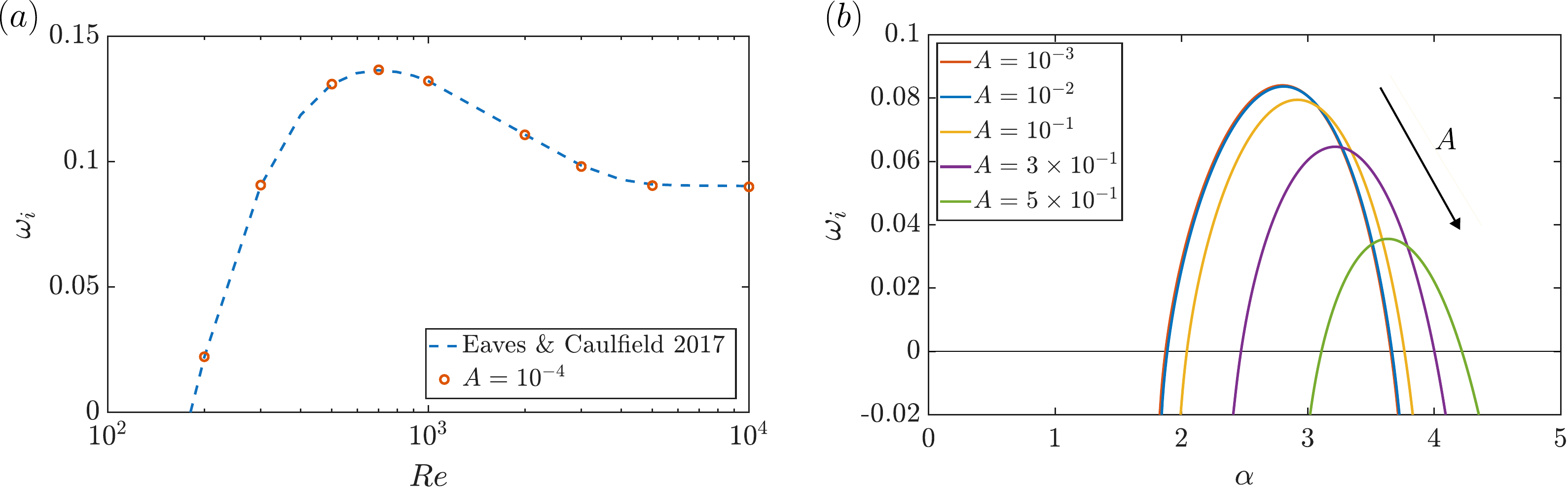}
    \captionsetup{justification=justified}
    \caption{(a) Maximum growth rate (calculated over varying Richardson numbers and disturbance wavenumbers) as a function of the Reynolds number, validating the proposed framework of section \ref{sec.linearEoSmodel} (red symbols) against the data from \cite{eaves2017multiple} (blue dashed line).     
    \rev{(b) Growth rate as a function of disturbance wavenumber for increasing values of $A$ at fixed $Ri=1$ and $Re=300$, using the constant-coefficient, linear EoS model of section \ref{sec.linearEoSmodel}.}}
    \label{fig.tripleDensityLayer}      
\end{figure}

The effect of the dimensionless number $A$ on the viscous Taylor-Caulfield instability is examined at $Ri=1$ (figure \ref{fig.tripleDensityLayer}b).
\rev{For verification purposes, the same calculations have been carried out with the general stability model of section \ref{sec.AEoSmodel}; the procedure and results are presented in appendix \ref{app.generalTCI}.}
In the previous section, the Boussinesq approximation was shown to be recovered when $A \ll 1$; this was expected from the discussion in section \ref{sec.constantcoeff}.
This Boussinesq regime is verified until approximately $A \simeq 10^{-2}$, as the growth rate is insensitive to $A$ for smaller values (figure \ref{fig.tripleDensityLayer}b).
Significant departures from these predictions are found from about $A=10^{-1}$ (which, to give an order of magnitude, approximately corresponds to a $30 K$ temperature difference in an air flow).
In this regime, non-Boussinesq effects become non-negligible and their effect is found stabilising as the growth rate quickly decreases with $A$.
The instability is also shifted towards higher wavenumbers.
These results are consistent with those of \cite{heifetz2015stratified}, obtained for an inviscid and alternative version of the TCI (with a discontinuous baseflow density profile and an unbounded domain).
Note that, in that work, non-Boussinesq effects were examined by varying the Froude number at a fixed Richardson number, which is equivalent to varying $A$, since $A=Ri Fr^2$.
\rev{The present study extends this work by accounting for viscous diffusion, which also allows for thermal expansion---this effect is absent in the inviscid regime where no thermal diffusion occurs (see equation \ref{eq.divU2}).}

\begin{figure}
    \centering
    \includegraphics[angle=-0,trim=0 0 0 0, clip,width=0.99\textwidth]{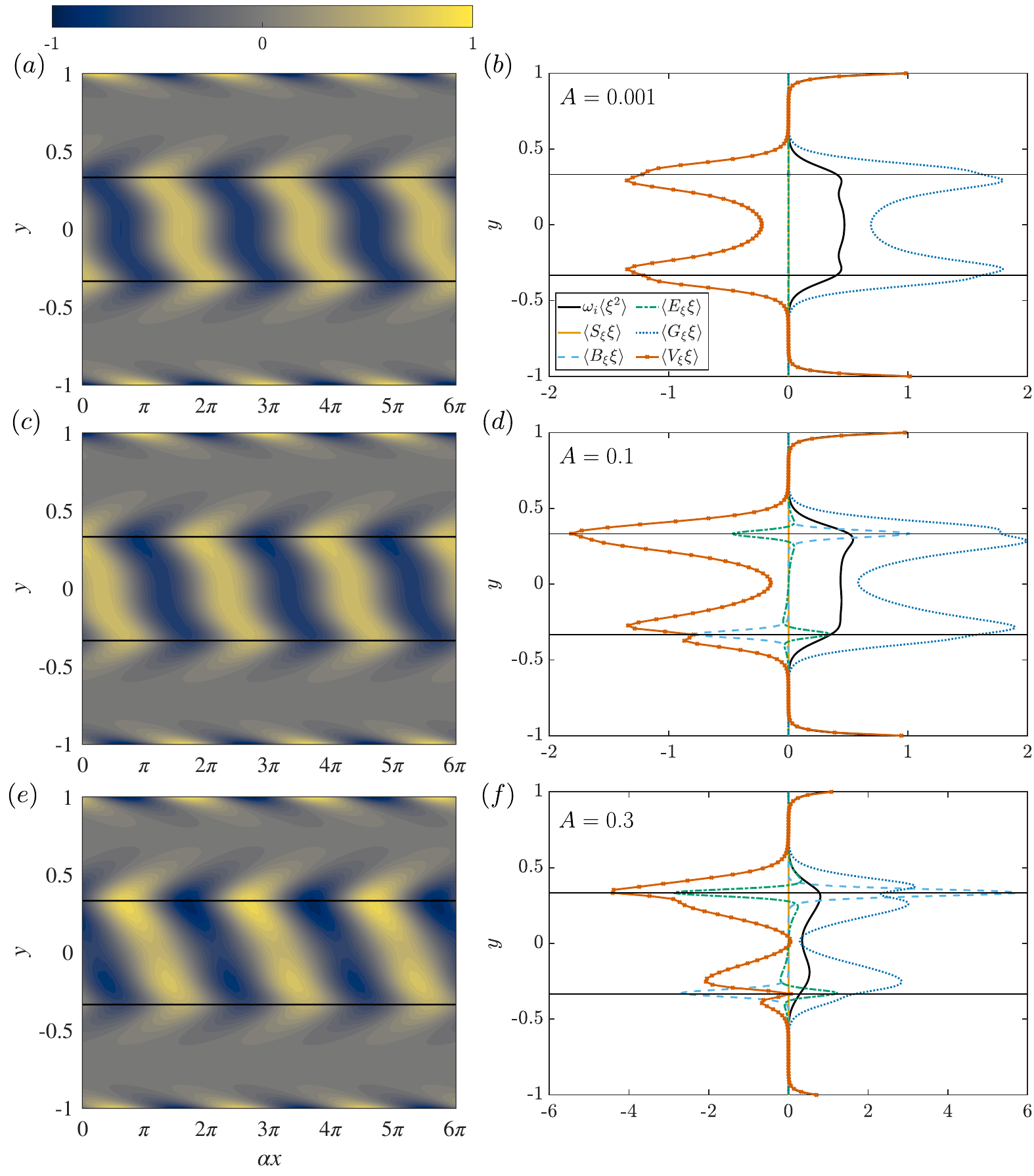}
    \captionsetup{justification=justified}
    \caption{\rev{TCI at $A=10^{-3} \> (a,b)$, $A=0.1 \> (c,d)$, and $A=0.3 \> (e,f)$.
    The other parameters are $Re=300$, $Ri=1$, $\alpha=3$.
    Left: vorticity field at an arbitrary time, normalised by its maximum value.   
    The solid horizontal lines indicate the location of the diffused density interfaces.
    Right: terms in the enstrophy equation, normalised, in each figure, by the maximum value reached by $\omega_i \langle \xi^2 \rangle$.}}
    \label{fig.TCI_fields}      
\end{figure}

The disturbance vorticity is now examined to gain further insight into the role of non-Boussinesq and viscous effects.
In the inviscid Boussinesq regime, the TCI can be understood from the interaction between two gravity waves, which are vorticity waves developing over density interfaces \citep{carpenter2011instability}.
In our study, these interfaces are diffused and located at $y=-1/3$ and $y=1/3$ (figure \ref{fig.sketch}b).
Figures \ref{fig.TCI_fields}(a,c,e) show the two-dimensional disturbance vorticity field at an arbitrary time, computed as $\Re \{ \xi(y) \exp \left[ i (\alpha x - \omega t ) \right] \}$, where $\Re$ denotes the real part of a complex quantity.
In the Boussinesq limit $A=10^{-3}$ (figure \ref{fig.TCI_fields}a), the disturbance vorticity exhibits local extrema along the two diffused interfaces: these are the interacting gravity waves, which are phase-locked.
This is in qualitative agreement with the viscous, Boussinesq calculations of \cite{eaves2017multiple}.
Two vorticity waves are also present at the bottom and top walls, respectively, as a result of viscous diffusion.
These two waves do not contribute to the vorticity dynamics of the TCI.
However, they can interact with gravity waves and give rise to what \cite{eaves2017multiple} termed the viscous Holmboe-wave instability (not studied here).
As $A$ in increased to and above 0.1 (where non-Boussinesq effects were previously shown to affect the instability growth rate), the vorticity field retain, overall, a structure similar to that of the Boussinesq limit (figures \ref{fig.TCI_fields}c,e).
This suggests that the gravity-wave interaction mechanism is still at play.
There are, however, significant differences: it can be noticed that some of the the symmetries are now broken.
The amplitude of the vorticity is now greater at the upper density interface than at the lower interface, and the location of the lower local maximum is now shifted slightly above at $y=-1/3$ .
This is better viewed by examining the profile of the vorticity magnitude, which can be assessed via $\xi^2$, the disturbance enstrophy.
An enstrophy equation can also be established to assess where and to what extent the vorticity sources identified in section \ref{sec.vorticity_eq} affect the vorticity magnitude.
The spatial average of any quantity $q$ is introduced as
\begin{align}\label{eq.avg}
  \langle q \rangle = \frac{1}{l_x}  \int_0^{l_x} q \mathrm{d}x ,
\end{align}
where $l_x = 2 \pi / \alpha$.
Multiplying the disturbance vorticity equation \eqref{eq.linear_vorticity} by $\xi$ and averaging according to equation\eqref{eq.avg}, the following enstrophy equation is established:
\begin{align} \label{eq.enstrophy}
  \Derp{\langle \xi^2/2 \rangle}{t} = \langle S_\xi \xi \rangle + \langle B_\xi \xi \rangle + \langle G_\xi \xi \rangle + \langle E_\xi \xi \rangle + \langle V_\xi \xi \rangle
\end{align}
In practice, the terms of this equation are computed by writing the vorticity equation \eqref{eq.linear_vorticity} in Fourier space, multiplying it by $\xi^*$ (the complex conjugate of the disturbance vorticity), and taking the real part. 
We note that $\derp{\langle \xi^2/2 \rangle}{t} = \omega_i \langle \xi^2 \rangle$, where $\omega_i$ is the growth rate of the mode, which therefore determines the sign of the left-hand side of equation \eqref{eq.enstrophy} (positive in the case of an instability).
Equation \eqref{eq.enstrophy} is a diagnostic equation that informs us about the contribution of each vorticity source, introduced in equation \eqref{eq.linear_vorticity}, to the magnitude of the disturbance vorticity.
The right-hand side terms are correlations between the sources and the vorticity.
They account for phase differences, and may therefore contribute either positively or negatively to the vorticity magnitude.

The terms of the enstrophy equation are shown in figures \ref{fig.TCI_fields}(b,d,f).
Only the buoyancy and viscous terms, $\langle G_\xi \xi \rangle$ and $\langle V_\xi \xi \rangle$, respectively, contribute to the disturbance enstrophy in the Boussinesq limit $A=10^{-3}$ (figure \ref{fig.TCI_fields}b)---baroclinic and thermal expansion effects are expected to be negligible in this regime.
The shear term is always zero for the TCI since the baseflow vorticity is constant.
The buoyancy contribution is positive and primarily localised around the density interfaces, consistent with the presence of gravity waves.
Viscosity contributes negatively in these regions as it diffuses the buoyancy-generated vorticity away from the interfaces.
As previously noted, the resulting enstrophy profile is symmetric at $A=10^{-3}$, but becomes skewed for $A=0.1$ and $A=0.3$ (figures \ref{fig.TCI_fields}d,f).
As $A$ increases, a growing positive contribution of the baroclinic torque $\langle B_\xi \xi \rangle$ is observed at the upper density interface.
Conversely, the baroclinic contribution is negative at the lower interface; it is also of smaller magnitude.
The buoyancy contribution remains positive and similar to that observed in the Boussinesq limit.
Altogether, this results in a larger-amplitude vorticity wave along the upper interface than along the lower interface.
This asymmetry caused by the baroclinic torque was noted in the theoretical analysis of \cite{heifetz2015stratified} in the inviscid, non-Boussinesq regime, and was shown to weaken the wave-interaction mechanism, thereby stabilising the TCI.
This is in agreement with the decreasing growth rate observed in figure \ref{fig.tripleDensityLayer}(b).
For $A=0.1$ and $A=0.3$ (non-Boussinesq regime), the vorticity of our system is furthermore affected by thermal expansion.
It is found to always oppose the baroclinic contribution, but with a significantly smaller magnitude (figures \ref{fig.TCI_fields}d,f).
Thus, diffusion-induced thermal expansion weakens the stabilising effect of the baroclinic torque on the TCI, since it reduces the asymmetry of the vorticity waves along the two interfaces.

%%%%%%%%%%%%%%%%%%%%%%%%%%%%%%%%%%%%%%%%%%%%%%%%%%%%%%%%%%%%%%%%%%%
%%%%%%%%%%%%%%%%%%%%%%%%%%%%%%%%%%%%%%%%%%%%%%%%%%%%%%%%%%%%%%%%%%%

\subsection{Supercritical fluid instability} \label{sec.LSTscf}
Fluids at supercritical pressures and temperatures exhibit steep property variations (in particular, viscosity and density) near the thermodynamic region known as the Widom line \citep{banuti2015crossing}.
The SCF instability, described in the introduction, is observed when a SCF flow develops a temperature gradient such that the Widom line lies within the flow domain \citep{ren2019boundary}.
We propose to apply the general framework presented in section \ref{sec.AEoSmodel} to a SCF flow in a differentially heated plane Couette flow. 
This setup is used first for validation purposes, recovering the inviscid SCF instability, and subsequently to gain new insights into the respective roles of viscosity and buoyancy.
We stress that, given the structure and magnitude of the density stratification, the viscous Boussinesq approximation is not applicable in this context.    

\subsubsection{Validation}\label{sec.scf_validation}
The general model of section \ref{sec.AEoSmodel} is first validated against an inviscid, non-buoyant calculation of \cite{bugeat2024inst}, which was obtained by solving the Rayleigh equation.
\rev{The van der Waals EoS is used to model the SCF and reads
\begin{align} \label{eq.vdw}
      \check{p} = \frac{8 \check{\rho} \check{T}}{3 - \check{\rho}} - 3 \check{\rho}^2,
  \end{align} 
\noindent where the (dimensionless) reduced variables $\check{p} = p^* / p_c^*$, $\check{\rho}=\rho^* / \rho_c^*$ and $\check{T}= T^* / T_c^*$ have been introduced.
Here, $p_c^*$, $T_c^*$ and $\rho_c^*$ are the critical pressure, temperature and density, respectively.
Two dimensionless parameters control the supercritical fluid: the ambient reduced pressure, $\check{p}_0 = p_0^* / p_c^*$, and the reduced temperature at the lower plate, $\check{T}_0 = T_0^* / T_c^*$.
The reduced density at the lower plate, $\check{\rho}_0 = \rho_0^* / \rho_c^* $ is deduced by numerically solving the non-linear equation \eqref{eq.vdw}.
The dimensionless function $\Phi$, which appears in the general model in equation \eqref{eq.phi1}, must be expressed with the dimensionless variables $T=T^*/T_0^*$ and $\rho=\rho^*/\rho_0^*$ (section \ref{sec.NL}) rather than with the above reduced variables.
Manipulating equation \eqref{eq.vdw}, we find that
\begin{align*}
  \Phi(\rho,T) = \left( \frac{3}{Z_0} \right) \frac{\rho T}{3 - \check{\rho}_0 \rho} - 3 \left( \frac{ \check{\rho}_0^2}{ \check{p}_0} \right) \rho^2 ,
\end{align*}
where we have introduced the compressibility factor at the reference thermodynamic point, $Z_0 = p_0^* / \rho_0^* r^* T_0^*$.
The compressible factor at the critical point is $Z_c=3/8$ for the van der Waals EoS.
It follows that $Z_0 = 3  \check{p}_0 /  8 \check{\rho}_0  \check{T}_0$.
}

\begin{figure}
    \centering
    \includegraphics[angle=-0,trim=0 0 0 0, clip,width=0.99\textwidth]{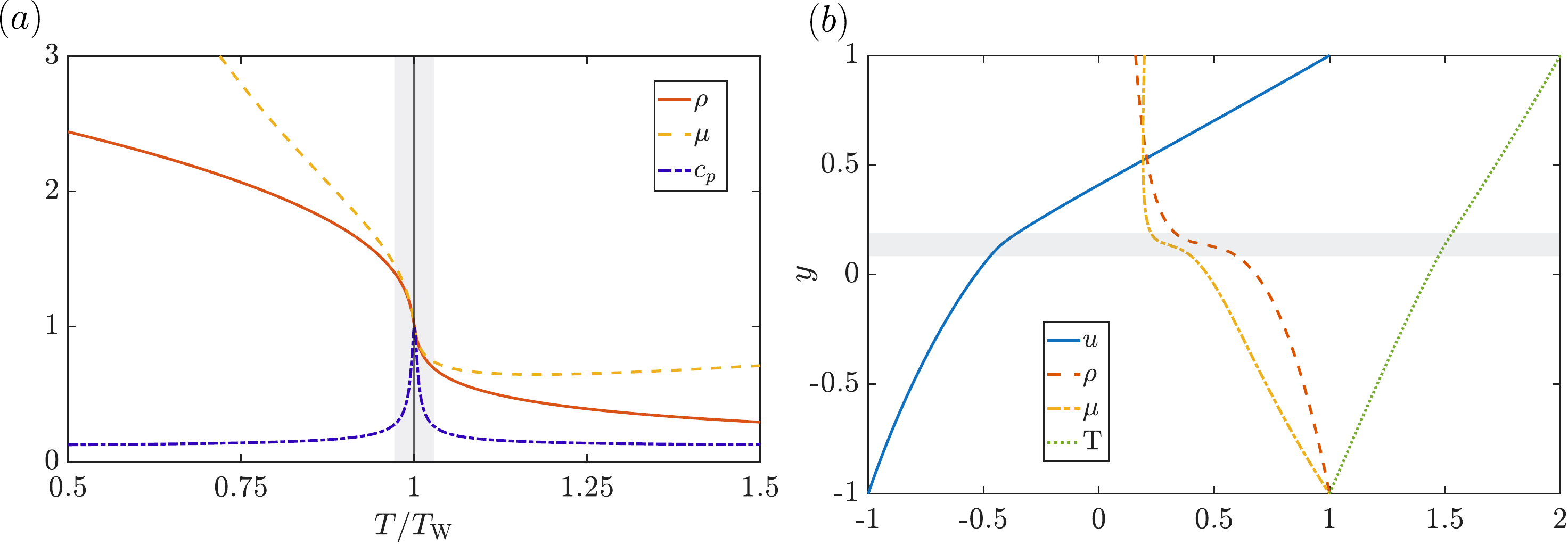}
    \captionsetup{justification=justified}
    \caption{\rev{(a) Density, dynamic viscosity, and heat capacity as a functions of temperature at a fixed supercritical reduced pressure $\check{p}_0 = 1.09$. 
    The temperature is normalised by the temperature at the Widom line, $T_\mathrm{W}$, and the fluid properties by their value at $T_\mathrm{W}$. 
    The black vertical line indicates $T/T_\mathrm{W}=1$.
    (b) Baseflow profile considered in the validation case ($\check{p}_0 = 1.
    09$, $\Lambda=1$).
    The grey shaded area highlights the region of strong fluid-property gradients.}}
    \label{fig.thermoVdW_and_baseflow}      
\end{figure}

\rev{We use the dynamic viscosity and thermal conductivity laws for SCFs proposed by \cite{jossi1962viscosity} and \cite{stiel1964thermal}, respectively; details can be found in appendix \ref{app.diff}.}
The heat capacity at constant volume $c_v$ is assumed constant \citep{johnston2014advances}; the calculation of $c_p$ from this assumption is shown in appendix \ref{app.cv}.
This involves the molecular complexity of the fluid species, which is an additional dimensionless physical parameter given by $C_{v,0}^*/R^*$, where $C_{v,0}^*$ the molar heat capacity at constant volume and $R^*$ the universal gas constant.
\rev{These fluid properties are plotted at constant pressure in figure \ref{fig.thermoVdW_and_baseflow}(a).
By definition, the Widom line is located where $c_p$ reaches a maximum, which can be identified at each supercritical pressure, thereby defining a `line' in pressure-temperature space.
Density and dynamic viscosity exhibit strong, localised gradients in the vicinity of the Widom line.}

\begin{figure}
    \centering
    \includegraphics[angle=-0,trim=0 0 0 0, clip,width=0.99\textwidth]{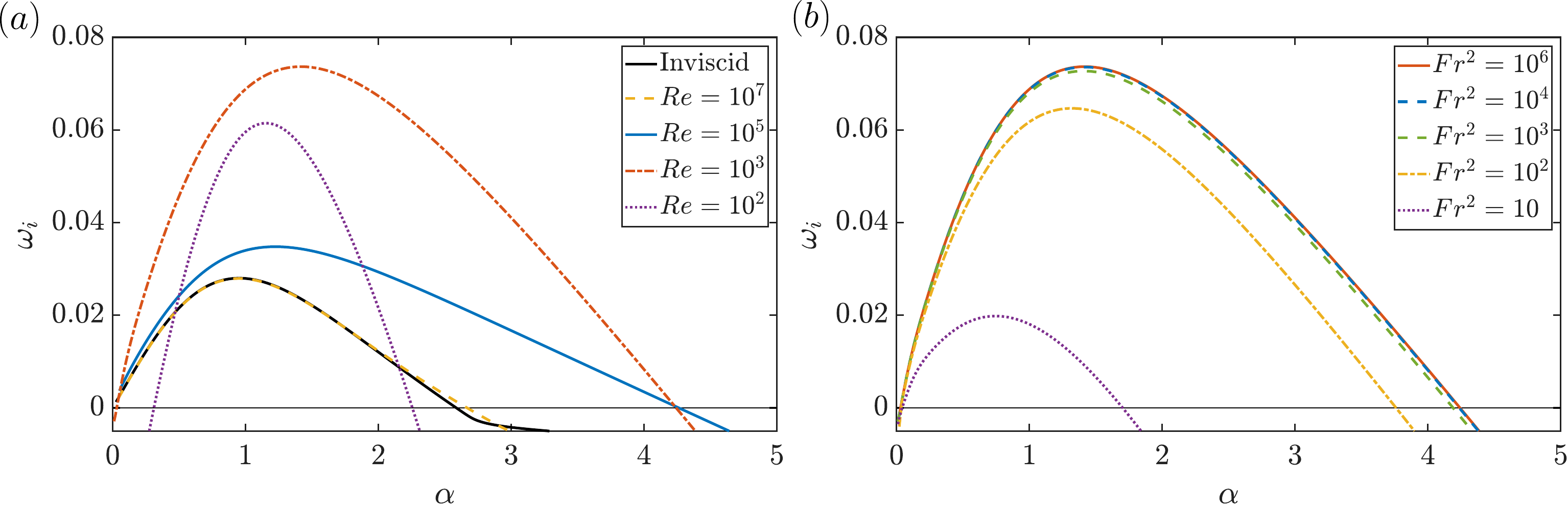}
    \captionsetup{justification=justified}
    \caption{Growth rate as a function of wavenumber for a van der Waals fluid at the supercritical reduced pressure $\check{p}_0=1.09$, with $\rev{\Lambda}=1$, $Pr=1$ and $C_{v,0}^*/R^*=5$. 
    (a) Effect of the Reynolds number at $Fr^2=10^6$. 
    The solid black line shows the non-buoyant, inviscid result from \cite{bugeat2024inst}. 
    (b) Effect of the Froude number at $Re=10^3$.}
    \label{fig.VdW}      
\end{figure}

\rev{The linear stability validation is carried out at $\check{p}_0 = 1.09$.
The baseflow temperatures at the bottom and top plates are chosen such that the Widom line lies around the centre line; more details on this procedure and the calculation of the baseflow can be found in \cite{bugeat2024inst}.
The resulting baseflow is shown in figure \ref{fig.thermoVdW_and_baseflow}(b). 
The signature of the Widom line is observed where the density and viscosity gradients are the strongest. 
Contrary to the constant-viscosity case, the baseflow shear varies and is smaller in regions where the baseflow viscosity is greater since the momentum equation verifies $\ubar' \propto 1/\mubar$.}
A large Froude number ($Fr^2=10^6$) is used to approach the non-buoyant regime, whilst $\rev{\Lambda} = 1$ and $C_{v,0}^*/R^*=5$. 
The inviscid, long-wave instability of the reference study is recovered at $Re=10^7$ (figure \ref{fig.VdW}a).
A small discrepancy remains near the instability cutoff, \rev{where viscous effects still moderately affect these higher wavenumbers. 
This is consistent with the fact that viscosity acts more strongly at smaller lengthscales \citep{charru2011hydrodynamic}.}

\subsubsection{Viscous and buoyant effects}
The effect of viscosity on the supercritical fluid instability is examined by  gradually reducing the Reynolds number from the inviscid, non-buoyant case examined in the previous subsection.
A non-monotonic behaviour is observed (figure \ref{fig.VdW}a).
Reducing $Re$ until $\mathcal{O}(10^3)$ increases the growth rate; viscosity has a destabilising effect.
As $Re$ is further reduced, the growth rate decreases and the flow eventually becomes stable for $Re=\mathcal{O}(10)$, where viscous dissipation dominates.
This behaviour was also observed in flat-plate boundary layer flows at supercritical pressure \citep{ren2019boundary,bugeat2022new}.
More generally, the occurrence of the destabilising effect of viscosity is well known in hydrodynamic instabilities---e.g. in TS waves \citep{charru2011hydrodynamic}.
However, in the present case, the long-wave SCF instability is intrinsically inviscid.
Possible scenarios accounting for this viscous destabilisation are discussed in the next subsection by examining the disturbance vorticity.

The effect of buoyancy is studied by setting $Re=10^3$ whilst decreasing the Froude number from the non-buoyant regime ($Fr^2=10^6$), which is verified until $Fr^2 \simeq 10^4$ (figure \ref{fig.VdW}b).
We observe appreciable buoyancy effects at $Fr^2 \simeq 10^3$, where a small but noticeable decrease of the growth rate is observed for the shorter wavelength.
As $Fr$ is further reduced, the growth rate is significantly reduced.
This stabilising behaviour of buoyancy in stably-stratified configurations is in agreement with the fully compressible (hence non-Boussinesq) calculations of \cite{boldini2026first}, who observed that TS waves in SCF boundary-layer flows were damped as the Richardson number increases.

\subsubsection{Vorticity dynamics} \label{sec.SCFI_vorticity}
\begin{figure}
    \centering
    \includegraphics[angle=-0,trim=0 0 0 0, clip,width=0.96\textwidth]{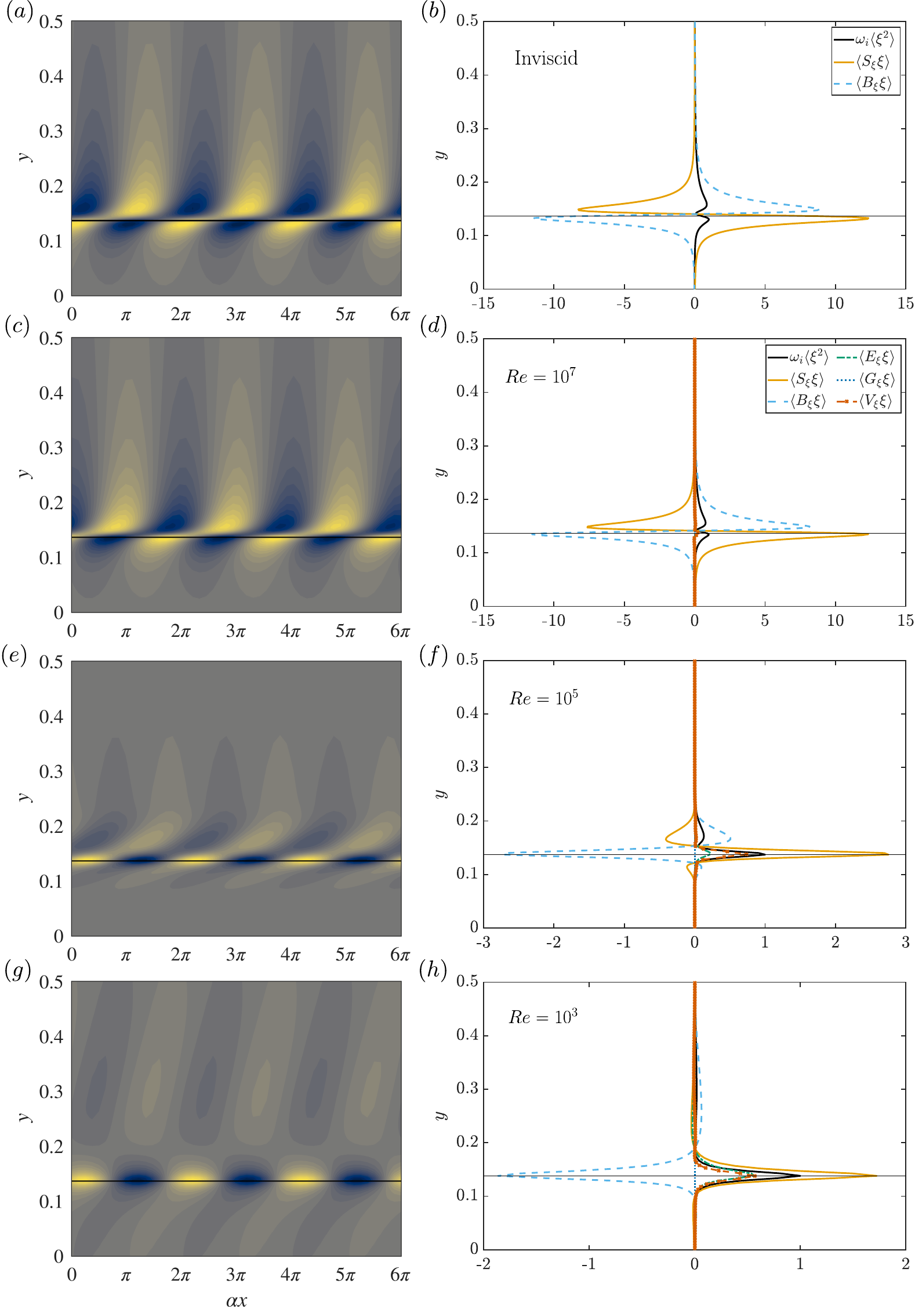}
    \captionsetup{justification=justified}
    \caption{\rev{SCF instability in the inviscid limit (a,b) and at $Re=10^7 \> (c,d)$, $Re=10^5 \> (e,f)$, $Re=10^3 \> (g,h)$.
    The inviscid results are obtained by solving the Rayleigh equation.
    Calculations are carried out at $\alpha=1$, $\check{p}_0=1.09$ and $Fr^2=10^6$.
    See figure \ref{fig.TCI_fields} for the description of the columns and the colour bar.
    The solid horizontal line indicates the Widom line.
    Note that only a portion of the numerical domain is shown in order to visualise the vorticity waves around the Widom line.}}
    \label{fig.SCFI_fields}      
\end{figure}

We focus our attention on the destabilisation of the SCF instability as the Reynolds number is gradually reduced from the inviscid regime to finite values.
We consider only the non-buoyant case $Fr^2=10^6$.
Figures \ref{fig.SCFI_fields}(a,c,e,g) show the disturbance vorticity fields in a portion of the domain that contains the Widom line.
Note that this excludes the unsteady boundary layers developing near the top and bottom walls.
In the inviscid regime, two phase-locked vorticity waves develop around the Widom line (figure \ref{fig.SCFI_fields}a).
An explanation of the physical origin of these waves, whose interaction gives rise to the SCF instability, was proposed by \cite{bugeat2024inst}; a condensed overview is provided below.
The baseflow contains sharp variations, or `diffused interfaces', in both the density and vorticity profiles around the Widom line (figure \ref{fig.thermoVdW_and_baseflow}b).
Two vorticity waves are therefore supported along these diffused interfaces, driven by the baroclinic torque and the vertical transport of baseflow vorticity, respectively.
The superposition (or interference) of these two waves gives rise to two lower-amplitude waves, as observed in figure \ref{fig.SCFI_fields}(a).
This is because the diffused density interface is located slightly above the diffused vorticity interface.
This inviscid scenario is confirmed by the positive contributions of the baroclinic term  $\langle B_\xi \xi \rangle$ and the shear term $\langle S_\xi \xi \rangle$ to the enstrophy above and below the Widom line, respectively (figure \ref{fig.SCFI_fields}b).
Conversely, $\langle B_\xi \xi \rangle$ becomes negative below the Widom line, whereas $\langle S_\xi \xi \rangle$ does so above it. 
\begin{figure}
    \centering
    \includegraphics[angle=-0,trim=0 0 0 0, clip,width=0.95\textwidth]{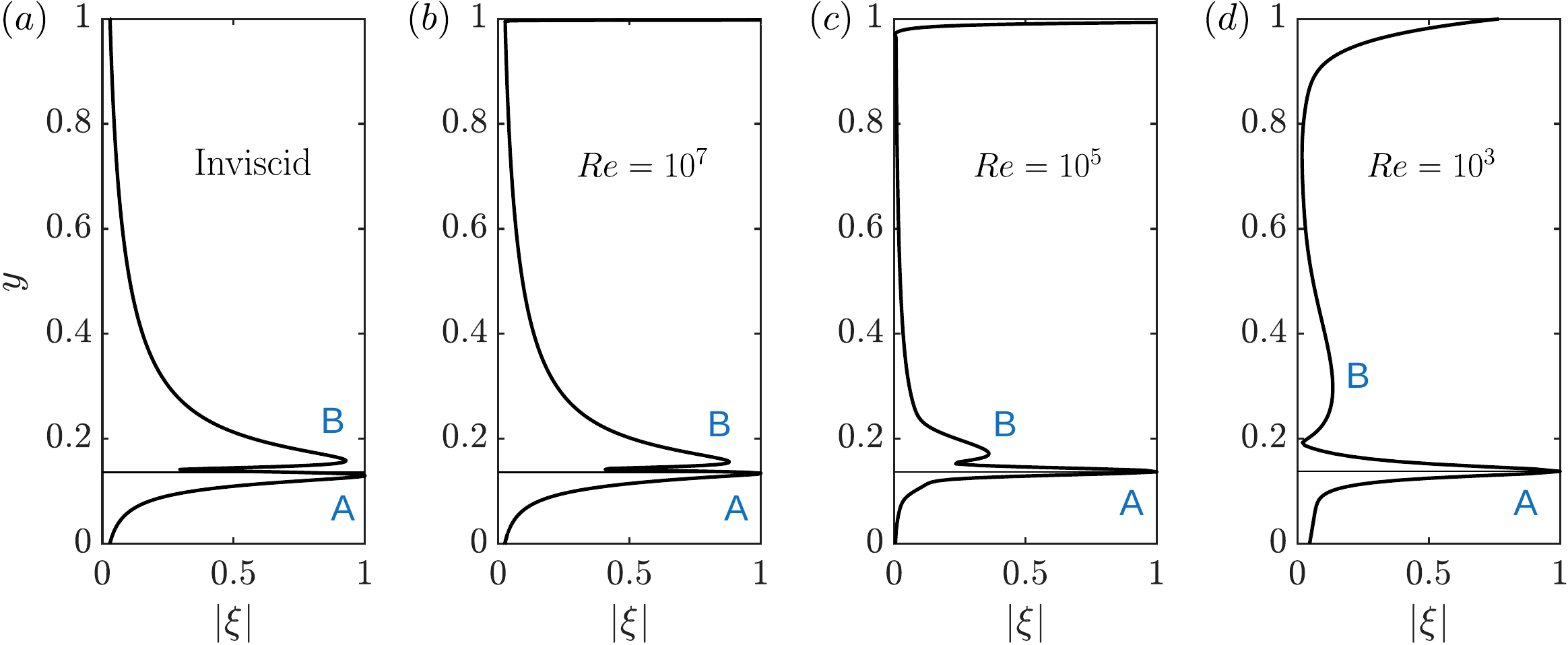}
    \captionsetup{justification=justified}
    \caption{\rev{Complex modulus of the disturbance vorticity $\xi$ .
    Parameters are the same as in figure \ref{fig.SCFI_fields}.
    The figure extends up to the top plate, showing the upper boundary layer.
    Labels A and B denote the two local maxima referred to in the main text.
    The solid horizontal line indicates the Widom line.}}
    \label{fig.SCFI_vorticity}      
\end{figure} 

The vorticity fields in the inviscid regime and at $Re=10^7$ are nearly identical (figure \ref{fig.SCFI_fields}a,c), as expected from the earlier validation of the growth rate.
As the Reynolds number is reduced to $10^5$, the vorticity field exhibits a significantly different structure (figure \ref{fig.SCFI_fields}e). 
The lower vorticity wave becomes dominant, whilst the upper vorticity wave has moved upwards.
At $Re=10^3$, this behaviour is even more pronounced.
% : the upper wave is much weaker and is shifted further up.
The distance between the two waves therefore continues to increase as the Reynolds number is reduced, although the waves remain phase-locked.
Figure \ref{fig.SCFI_vorticity} provides an alternative visualisation through the profile of the complex modulus of the disturbance vorticity.
As the Reynolds number is reduced, a growing asymmetry develops between the two waves (labelled A and B).
The spatial support of wave B increases (the wave is less localised) whilst this wave shifts upwards away from the Widom line.

The enstrophy equation can be used to examine these viscosity-induced reorganisations of the vorticity field.
The baroclinic and shear terms, ${\langle B_\xi \xi \rangle}$ and $\langle S_\xi \xi \rangle$, are significantly altered at $Re=10^5$ and $Re=10^3$ (figures \ref{fig.SCFI_fields}f,h) compared with those in the inviscid and high-Reynolds-number regimes (figures \ref{fig.SCFI_fields}b,d).
Both terms retain two local extrema of opposite signs, but the magnitudes and locations of these peaks exhibit behaviour similar to that observed for the vorticity in figure \ref{fig.SCFI_vorticity}.
This suggests that the two phase-locked vorticity waves are still primarily generated by the baroclinic--shear mechanism described above for the inviscid regime, although they undergo significant changes.
These changes can be seen as an indirect consequence of viscosity, which affects all flow variables and hence the magnitude and distribution of $B_\xi$ and $S_\xi$, but does not explicitly appear in these source terms (equation \ref{eq.linear_vorticity}).

More directly, finite Reynolds numbers allow for thermal diffusion, which in turn allows for thermal expansion, as noted, for example, in section \ref{sec.TCI_varyA}.
Significant density variations are expected where $\rhoT$ is large (equation \ref{eq.rhoT1}).
Consistent with this, the contribution of thermal expansion to the enstrophy, $\langle E_\xi \xi \rangle$, is localised around the Widom line (figures \ref{fig.SCFI_fields}f,g), where the density gradients are largest.
This contribution is positive.
The same observation is made for the `direct' viscous effect associated with the viscous stresses, whose positive contribution to the vorticity magnitude via the term $\langle V_\xi \xi \rangle$ is localised around the Widom line.
Overall, these two diffusion-related effects, whose relative importance increases as $Re$ is reduced, reinforce the lower vorticity wave (A) whilst having no appreciable effect on the upper wave (B).

We now consider an additional vorticity wave located away from the Widom line.
Figure \ref{fig.SCFI_vorticity} extends to the top of the numerical domain ($y=1$), showing the presence of a vorticity wave near the upper wall; this wave corresponds to the disturbance boundary layer (a similar wave develops near the bottom plate),
The thickness of this layer, which results from unsteady viscous diffusion of vorticity from the wall, is expected to scale as $(\nu / \omega_r)^{1/2}$.
Assuming that $\omega_r / \alpha = \mathcal{O}(1)$ gives the dimensionless scaling $(\alpha Re)^{-1/2}$.
Note that $\alpha$ is fixed and is equal to 1 in this section.
At $Re=10^7$ and $Re=10^5$, the boundary layer is very thin (figures \ref{fig.SCFI_vorticity}b,c).
At $Re=10^3$, however, it penetrates significantly into the domain (figure \ref{fig.SCFI_vorticity}d), raising the possibility of an interaction with wave B; this is discussed in the next subsection.
As a side note, given the scaling $(\alpha Re)^{-1/2}$, vorticity diffusion from the wall is expected to become dominant at low wave numbers, which accounts for the stabilisation of long waves (figure \ref{fig.VdW}); the lower the Reynolds number, the higher the low-wavenumber cutoff.
Note that there is no boundary layer in the inviscid case (figure \ref{fig.SCFI_vorticity}a) and all long waves are unstable, as  shown theoretically by \cite{bugeat2024inst}.

%%%%%%%%%%%%%%%%%%%%%%%%%%%%%%
\subsubsection{Discussion}
The observations made regarding disturbance vorticity dynamics do not provide an immediate explanation for the viscous destabilisation phenomenon.
The diagnostic enstrophy equation shows that baroclinic torque, the transport of baseflow vorticity transport, thermal expansion, and viscous stresses all contribute significantly to the production or destruction of disturbance vorticity.
From a modelling perspective, this highlights the importance of accounting for these effects and supports the use of the proposed model.
From a physical viewpoint, it raises questions about the instability mechanism at play at finite Reynolds number.
Three scenarios may be envisaged based on the observations in section \ref{sec.SCFI_vorticity}.

(i) It was shown that the inviscid mechanism at the origin of the two interacting vorticity waves persist in the viscous regime, albeit with substantial changes in the amplitudes and locations of the two waves.
Diffusion-induced effects (thermal expansion and viscous stresses) were shown to further amplify the lower vorticity wave (A).
Because these effects become stronger as the Reynolds number is reduced, it can be argued that they are responsible for the destabilisation of the inviscid instability.
However, in the meantime, the distance between the waves grows and an asymmetry develops in their amplitude, which weakens the interaction, as seen for the TCI \citep{heifetz2015stratified}.
Because these two qualitative effects oppose each other, it is not possible to draw a conclusion without a theoretical analysis of the viscous wave-interaction problem.

(ii) The observation of a vorticity wave developing along the wall, corresponding to the unsteady boundary layer, suggests that this wave may favourably interact with the central, Widom-line-induced wave B.
This scenario is similar to that proposed by \citep{eaves2017multiple} to explain the viscous Holmboe-wave instability, which is interpreted as the result of an interaction between a gravity wave and a viscous wall wave.
At $Re=10^3$, such an interaction may occur as the boundary layer grows significantly towards the Widom line whilst the vorticity wave B shifts towards the wall (figure \ref{fig.SCFI_vorticity}d).
However, this scenario is less plausible at $Re=10^5$ as the two waves appear isolated from each other (figure \ref{fig.SCFI_vorticity}c).
A detailed analysis would be required to assess the strength of this interaction and the conditions for it to occur.

(iii) If we now assume that all wave interactions only gradually weaken as the Reynolds number is reduced, an entirely different instability mechanism would have to emerge at finite Reynolds numbers, superseding the wave interaction.
By introducing new phase shifts between $u$ and $v$, viscous effects can increase the transfer of energy from the baseflow to the perturbation, represented by the production term ${\mathcal{P} = - \ubar \langle uv \rangle}$ \citep{schmid2002stability}.
This is how unstable viscous TS waves are usually understood: in the presence of viscosity, the no-slip condition at the wall breaks the inviscid phase quadrature between $u$ and $v$, allowing for $\mathcal{P}$ to overcome viscous dissipation \citep{charru2011hydrodynamic}.
Note that this energy perspective may obscure physically interpretable phenomena, such as baroclinic torque, because $\mathcal{P}$ is the only term capable of transferring energy to the perturbations; the other terms are either dissipative or redistributive.
Theoretical progress would then require the cause of this favourable phase shift to be clearly identified, thereby providing useful insights into mechanism driving the viscous destabilisation.

%%%%%%%%%%%%%%%%%%%%%%%%%%%%%%%%%%%%%%%%%%%%%%%%%%%%%%%%%%%%%%%%%%%%%%%%%%%%%
%%%%%%%%%%%%%%%%%%%%%%%%%%%%%%%%%%%%%%%%%%%%%%%%%%%%%%%%%%%%%%%%%%%%%%%%%%%%%
\section{Conclusion}
A dimensionless mathematical framework has been developed to analyse the linear stability of viscous stratified flows for fluids with arbitrary equations of state.
The formulation is of practical interest, as it requires only an equation of state of the form $p=\Phi(T,\rho)$ or $\rho=\phi(T)$.
The structure and magnitude of flow stratifications (e.g. density, viscosity) are unrestricted, and non-Boussinesq effects are accounted for.
The model is based on the low-Mach approximation, which filters out acoustic modes.
Overall, this framework is more versatile than the viscous Boussinesq approximation whilst removing the unnecessary complexity of fully compressible models at low Mach number. 
The general linear formulation can be found in equations (\ref{eq.L1}-\ref{eq.L5}), with a special case corresponding to a linear equation of state with constant coefficients provided in (\ref{eq.L1const}-\ref{eq.L5const}).

Two application cases have been considered to validate the framework and demonstrate its potential.
First, the Taylor-Caulfield instability is successfully recovered.
Non-Boussinesq effects are shown to stabilise this mode as the dimensionless number $A=\alpha^* \Delta T^*$ exceeds approximately $10^{-1}$.
This indicates that models based on the Boussinesq approximation are not valid in this regime.
Second, the supercritical fluid instability is validated in the inviscid, non-buoyant limit.
This flow cannot typically be modelled within the Boussinesq approximation, as relative density variations are of order unity and baroclinic effects, arising from inertial density variations, are essential to the instability mechanism.
Viscous effects are shown to further destabilise the flow over a range of Reynolds numbers, whilst buoyancy associated with stable stratification is stabilising.
\rev{The disturbance vorticity was shown to be affected by viscosity in a non-trivial manner, as it appears to both reinforce and weaken the existing inviscid mechanism (from a qualitative point of view).
Although the analysis remains inconclusive, several scenarios were suggested to explain the viscous destabilisation.}
These new results open several avenues for further investigation, and the proposed framework provides a practical, flexible tool to support future work.

\bigskip
\noindent \bb{Acknowledgements}
This work was funded by the EPSRC (project reference: UKRI1119). 
The author is grateful to the reviewers for their valuable suggestions, which have considerably improved the rigour, clarity, and overall quality of the original manuscript. In particular, Reviewer 2 suggested interpreting $\left( M / Fr \right)^2 \ll 1$ in terms of characteristic length scales.

\bigskip
\noindent \bb{Declaration of Interests} The author reports no conflict of interest.

\appendix
\section{Implicit equation of state} \label{app.imp}
The present framework can still be used if the EoS is an implicit function of the form
\begin{align}\label{eq.implicitEoS}
  \rev{\Pi} (p,T,\rho) = 0 .
\end{align}
Our linear model relies on the calculation of $\phiT$ and $\phirho$, which can be obtained by differentiating \eqref{eq.implicitEoS} by $T$ at constant $\rho$ and by $\rho$ at constant $T$, respectively.
We find
\begin{align}\label{eq.implicitDerivatives}
  \phiT  = \rev{ - \frac{ \partial_T \Pi }{\partial_p f}  \> \> , \> \> \phirho  = - \frac{ \partial_\rho \Pi }{\partial_p f} }
\end{align}

\rev{\section{Values of $\Psi$ and $\tilde{A}$}\label{app.Psi_Atilde}
The values of the dimensionless numbers $\Psi$ and $\tilde{A}$, introduced in section \ref{sec.NL}, are reported in table \ref{tab.values}.
These values were computed at a given pressure and temperature for air, water and supercritical \COt using the thermophysical property data from https://webbook.nist.gov/chemistry/fluid/ and www.engineeringtoolbox.com.
\begin{table}
     \begin{center}
     \def~{\hphantom{0}}
     \begin{tabular}{lllll}
         Fluid \hspace{2cm} & Pressure $p_0^*$ (bar) \hspace{0.3cm}  & Temperature $T_0^*$ ($K$) \hspace{0.3cm}  &  $\Psi$  \hspace{1.5cm} & $\tilde{A}$ \hspace{0.1cm}\\[5pt]
         Air                       & 1   & 293 &  0.29 &  1.0  \\
         Water                      & 1   & 293 &  $8.2 \times 10^{-5}$ &  $6.1 \times 10^{-2}$  \\
         Supercritical \COt       & 80  & 307 &  $5.5 \times 10^{-3}$ &  3.7  \\
     \end{tabular}
     \captionsetup{justification=justified}
     \caption{Value of the dimensionless numbers $\Psi$ and $\tilde{A}$ for three fluids.}
     \label{tab.values}
     \end{center}
    \end{table}
}

\section{Implementation of the TCI in the general model} \label{app.generalTCI}
\begin{figure}
    \centering
    \includegraphics[angle=-0,trim=0 0 0 0, clip,width=0.5\textwidth]{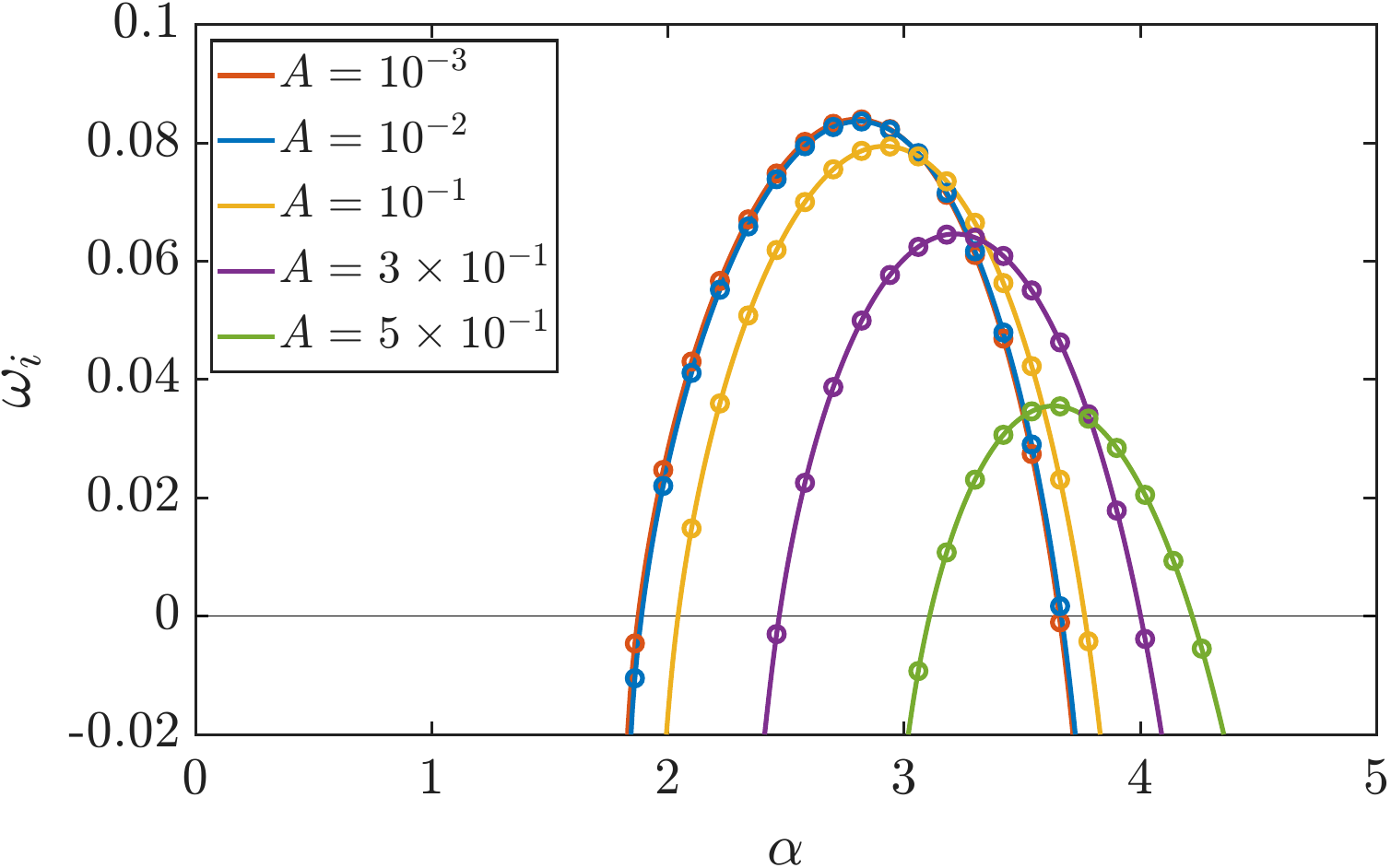}
    \captionsetup{justification=justified}
    \caption{Growth rate of the TCI as a function of the disturbance wavenumber with increasing $A$ at fixed $Ri=1$ and $Re=300$, using the constant-coefficient, linear EoS model of section \ref{sec.linearEoSmodel} (solid lines) and the general model of section \ref{sec.AEoSmodel} (symbols). 
    The consistency between the parameters of the two models is achieved via $Ri Fr^2 = A$.}
    \label{fig.verif}      
\end{figure}

The TCI was examined in section \ref{sec.TCI} using the constant-coefficient, linear-EoS model presented in section \ref{sec.constantcoeff}.
The general model for arbitrary equations of state developed (section \ref{sec.AEoSmodel}) can also be used to compute this instability, providing a useful verification of the stability code.
We use an EoS of the form $\rho = \phi(T)$ (see section \ref{sec.eosPhi}) and will assume that $\mu$, $\lambda$ and $c_p$ are constant.
A linear relationship between $\rho$ and $T$ is introduced as
\begin{align}\label{eq.EoSeaves}
  \phi(T) = 1 - \frac{A}{\rev{\Lambda}}  (T - 1) ,
\end{align}
along with a baseflow temperature profile
\begin{align}\label{eq.Teaves}
  \Tbar = 1 + \frac{\rev{\Lambda}}{2}  \left\{ 1 +  \frac{1}{2}  \left[  \tanh R \left( y - \frac{1}{3} \right) + \tanh R \left( y + \frac{1}{3} \right)   \right]  \right\}
\end{align}
which, when combined together, yields equation \eqref{eq.rhoEaves}.
Note that $T$ is made dimensionless by $T_0^*$ in this framework; this is why $\rev{\Lambda}$ appears in the above equations.
The Froude number is set so as to work with the same Richardson number as in the previous subsection, using $Fr^2 = A / Ri$. 
As expected, we recover the results obtained with the constant-coefficient, linear EoS model presented in section \ref{sec.TCI_varyA}, with calculations carried out for several Froude numbers at $Re=300$ (figure \ref{fig.verif}).

\rev{\section{Mesh convergence} \label{app.meshconv}
The independence of the results with respect to the number of mesh points $N$ is demonstrated in figure \ref{fig.meshconv} for both the TCI and SCF instability.
Higher Reynolds numbers classically require a larger number of mesh points to resolve the stronger gradients arising from weaker diffusion effects.
\begin{figure}
    \centering
    \includegraphics[angle=-0,trim=0 0 0 0, clip,width=0.995\textwidth]{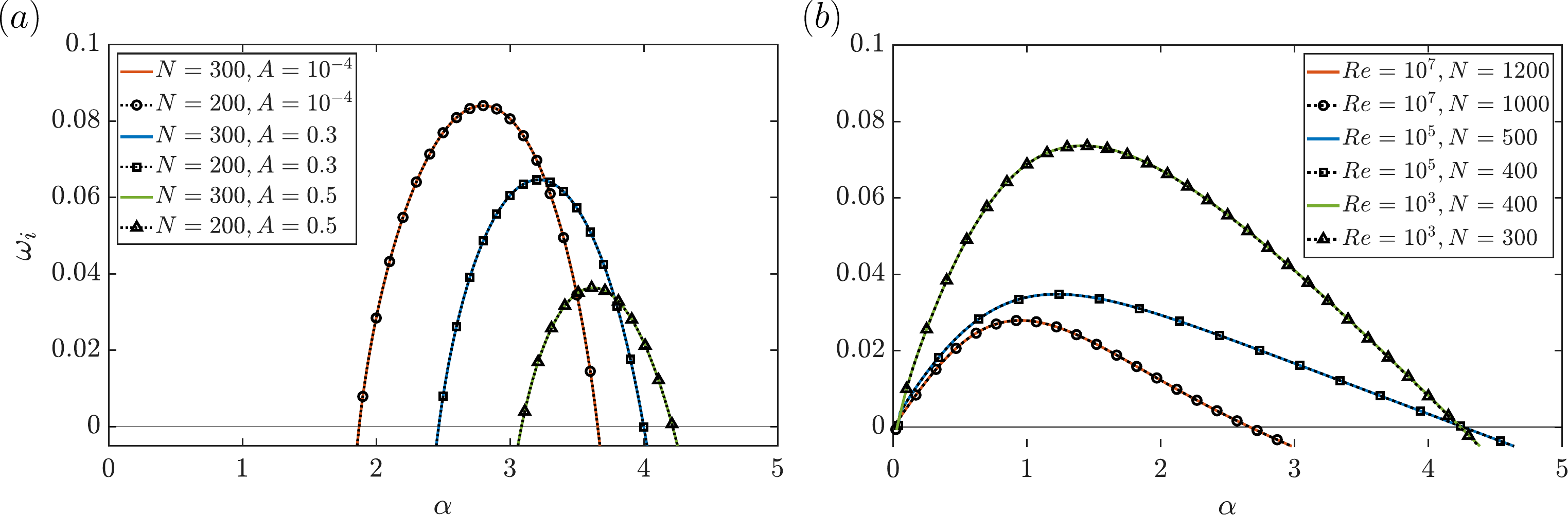}
    \captionsetup{justification=justified}
    \caption{\rev{Effect of $N$ on the computed growth rate.
    (a) TCI for different values of $A$ at $Ri=1$ and $Re=300$.
    (b) SCF instability for different values of $Re$ at $Fr^2=10^6$, $\check{p}_0 = 1.09$, and $\Lambda=1$.}}
    \label{fig.meshconv}      
\end{figure} 
}

\section{Note on heat capacities} \label{app.cv}
The formulation presented in section \ref{sec.general_case_summary_of_the_set_of_equations} assumes that a caloric model on the specific heat capacity at constant pressure, $c_p$, is available.
It is sometimes more convenient to work with a model on $c_v$.
\rev{For example, in section \ref{sec.scf_validation}, a van der Waals fluid is considered, for which $c_v$ is assumed constant \citep{johnston2014advances}.
In this case, $c_p$ is deduced from both $c_v$ and the EoS via the dimensional relation}
\begin{align}
  c_p^* =  c_v^*  + \frac{T^* (\phiT^*)^2}{(\rho^*)^2 \phirho^*} 
\end{align}
The dimensionless version of this equation when $c_v = c_{v,0}$ is constant is
\begin{align}\label{eq.cp}
  c_p =  \frac{1}{\gamma_0} + \Psi \frac{T (\phiT)^2}{\rho^2 \phirho} .
\end{align}
where $\gamma_0$ is the heat capacity ratio at $(p_0^*,T_0^*)$, and where $\Psi$ was first introduced in equation \eqref{eq.NDnumer}.
In the case of a gas or a supercritical fluid, we note that $\Psi$ can be rewritten,  as
\begin{align}
  \Psi =  \frac{Z_0}{\gamma_0 C_{v,0}^*/R^*} ,
\end{align}
where the usual dimensionless thermodynamic parameters appear: the compressibility factor $Z_0$ at $(p_0^*,T_0^*)$ and the molecular complexity of the species $C_{v,0}^*/R^*$, with $C_{v,0}^*$ the molar isochoric heat capacity and $R^*$ the universal gas constant.
This approach is adopted in section \ref{sec.LSTscf} with the van der Waals EoS.
We note the following relation with $\gamma_0$:
\begin{align}
  \gamma_0 =  1 + \frac{Z_0}{C_{v,0}^*/R^*} \frac{(\phiT_0)^2}{\phirho_0} ,
\end{align}
where we used $c_p = 1$ at $T=1$ and $\rho=1$ in equation \eqref{eq.cp}.

\section{Viscosity and thermal conductivity for SCFs} \label{app.diff}
\rev{The dynamic viscosity law established by \cite{jossi1962viscosity} for pure, non-polar supercritical fluids is used.
Using the dimensionless quantities defined in sections \ref{sec.NL} and \ref{sec.scf_validation}, the viscosity reads
\begin{align}
  \mu (T,\rho) =  \frac{m_1(T \check{T}_0) + m_2(\rho \check{\rho}_0)}{m_1(\check{T}_0) + m_2(\check{\rho}_0)} ,
\end{align}
where the functions $m_1$ and $m_2$ are defined as
\begin{align}
  m_1(x) &= 34 x^{0.94} \times 10^{-5} , \\
  m_2(x) &= 10^{-4} + \left( a_0 + a_1 x + a_2 x^2 + a_3 x^3 + a_4 x^4 \right)^4 ,
\end{align}
and where
\begin{align}
    a_0 &=  0.10230  , \\
    a_1 &=  0.023364 , \\
    a_2 &=  0.058533 , \\
    a_3 &= -0.040758 , \\
    a_4 &=  0.0093324 .
\end{align}}
\rev{In addition, the thermal conductivity law proposed by \cite{stiel1964thermal} for pure, non polar supercritical fluids is considered.
The dimensionless thermal conductivity defined in this paper then reads
\begin{align}
    \lambda (T,\rho) = \frac{0.07451 ( 0.307C_v/R + 0.539 ) l_1(T \check{T}_0) + l_2(\rho \check{\rho}_0)/Z_c^5}{0.07451 ( 0.307C_v^*/R^* + 0.539 ) l_1( \check{T}_0 ) + l_2( \check{\rho}_0 )/Z_c^5} ,
\end{align}
where the functions $l_1$ and $l_2$ are defined as
\begin{align}
  l_1(x) &= m_1(x) , \\ 
  l_2(x) &= 13.1 (e^{0.67x} - 1.069) \times 10^{-8} ,
\end{align}
and where $Z_c$ is the compressibility factor at the critical point ($Z_c=3/8$ for the van der Waals equation of state).
}

\bibliographystyle{jfm}
\bibliography{bibliSupercritical}

\end{document}